\documentclass[fleqn,tightenlines,preprint,preprintnumbers,letterpaper,superscriptaddress,nofootinbib,aps,eqsecnum]{revtex4-1}

\usepackage{amsmath}
\usepackage{amssymb}
\usepackage{wasysym}
\usepackage{bm}
\usepackage{latexsym}
\usepackage{stmaryrd}
\usepackage{textcomp}
\usepackage{verbatim}
\usepackage{slashed}
\usepackage{array}
\usepackage[dvips]{graphicx}
\usepackage{layout}
\usepackage{axodraw4j}
\usepackage{color}
\usepackage{cancel}
\usepackage[dvips]{hyperref}

\renewcommand{\arraystretch}{1.5}

\newcommand{\lp}{\left(}
\newcommand{\rp}{\right)}
\newcommand{\lb}{\left[}
\newcommand{\rb}{\right]}
\newcommand{\lc}{\left\{}
\newcommand{\rc}{\right\}}
\newcommand{\abs}[1]{\left|#1\right|}
\newcommand{\vareps}{\varepsilon}
\newcommand{\of}[1]{\left(#1\right)}
\newcommand{\eqnref}[1]{eq.~$\left(\ref{#1}\right)$}
\newcommand{\figref}[1]{fig.~\ref{#1}}
\newcommand{\secref}[1]{section~\ref{#1}}
\newcommand{\appref}[1]{appendix~\ref{#1}}
\newcommand{\tabref}[1]{table~\ref{#1}}

\newcommand{\Tabref}[1]{Table~\ref{#1}}

\newcommand{\etau}{$e\rightarrow\tau$}
\newcommand{\taueg}{$\tau\rightarrow e\gamma$}
\newcommand{\mueg}{$\mu\rightarrow e\gamma$}
\newcommand{\mue}{$\mu\rightarrow e$}
\newcommand{\mueee}{$\mu\rightarrow 3e$}
\newcommand{\taueee}{$\tau\rightarrow 3e$}
\newcommand{\emu}{$e\rightarrow\mu$}
\newcommand{\lqratio}{$\lambda_{1\alpha}\lambda_{3\beta}/M_{LQ}^2$}
\newcommand{\lqratiod}{$\lambda_{1\alpha}\lambda_{3\alpha}/M_{LQ}^2$}
\newcommand{\lfvet}{LFV(1,3)}
\newcommand{\lfvem}{LFV(1,2)}
\newcommand{\ifb}{${fb}^{-1}$}
\newcommand{\gev}{$GeV$}
\newcommand{\stilde}{$\tilde{S}_{1/2}^L$}

\begin{document}

\title{\Large Electron-to-Tau Lepton Flavor Violation at the Electron-Ion Collider}
\vspace{4.0cm}
\author{Matthew~Gonderinger}
\email{gonderinger@wisc.edu}
\affiliation{
{\small Department of Physics, University of Wisconsin-Madison} \\
{\small Madison, WI 53706, USA}}
\author{Michael~J.~Ramsey-Musolf}
\email{mjrm@physics.wisc.edu}
\affiliation{
{\small Department of Physics, University of Wisconsin-Madison} \\
{\small Madison, WI 53706, USA}}
\affiliation{
{\small Kellogg Radiation Laboratory, California Institute of Technology} \\
{\small Pasadena, CA, 91125, USA}}

\begin{abstract}

We analyze the potential sensitivity of a search for \etau\ conversion  at a proposed electron-ion collider (EIC) facility.  To that end, we calculate the cross sections for \etau\ events in a leptoquark framework assuming that the leptoquark masses are on the order of several hundred \gev\ or more.  Given present limits on leptoquarks from direct searches at HERA and rare decay processes, an EIC sensitive to 0.1~$fb$ \etau\ cross sections could probe previously unexplored regions of parameter space for these lepton flavor violating events (assuming 90~\gev\ center-of-mass energy and 10~\ifb\ integrated luminosity).  Depending on the species of leptoquark and flavor structure of the couplings, an EIC search could surpass the HERA and rare process sensitivity to \etau\ conversion amplitudes by as much as an order of magnitude or more.
We also derive updated limits on quark flavor-diagonal LFV leptoquark interactions using the most recent BaBar \taueg\ search. We find that limits from an EIC \etau\ search could be competitive with the most recent \taueg\ limit for a subset of the quark flavor-diagonal leptoquark couplings.  Using an SU(5) GUT model in which leptoquark couplings are constrained by the neutrino masses and mixing,  we illustrate how observable leptoquark-induced \etau\ conversion can be consistent with stringent LFV limits imposed by \mueg\ and \mue\ conversion searches. 
\end{abstract}

\preprint{NPAC-10-09}

\maketitle

\section{Introduction}\label{sec:intro}

The possibility that flavor may not be conserved by charged leptons at an observable level continues to be a topic of substantial interest in particle physics.  Although neutrino oscillations imply that charged lepton flavor violating processes such as \mueg\ are allowed in the Standard Model (SM), their rates are extraordinarily small --- and therefore unobservable --- because they are suppressed by the masses of the neutrinos.  Many models of physics beyond the SM predict rates for charged lepton flavor violation (LFV) processes that are both larger than the SM predictions and within reach of present or future experiments.  Hence, LFV is considered an important probe of new physics models such as SUSY $SO(10)$ grand unification theories (GUTs) (for a concise review of several models, see \cite{Albright:2008ke}), a supersymmetric SM with right-handed neutrinos \cite{Hisano:1995cp}, left-right symmetric models \cite{Cirigliano:2004mv}, and the Randall-Sundrum model (see for example \cite{Chen:2008qg} and the references therein).  LFV is also present in models containing leptoquarks with trans-generational couplings to leptons and quarks.  Various types of leptoquarks are predicted to exist, for example, in the Pati-Salam $SU(4)$ color model \cite{Pati:1974yy} and non-supersymmetric GUTs utilizing $SU(5)$ \cite{Georgi:1974sy} and $S0(10)$ \cite{Fritzsch:1974nn} symmetry groups.  

Extensive searches for charged lepton flavor violation between the first and second lepton generations (which we will refer to as ``\lfvem '' for brevity) have placed stringent experimental limits on processes such as \mue\ conversion (from the SINDRUM II collaboration \cite{Bertl:2006up_SINDRUM}), \mueg\ (from the MEGA collaboration \cite{Brooks:1999pu_MEGA,Ahmed:2001eh}), and \mueee\ (from the SINDRUM collaboration \cite{Bellgardt:1987du_SINDRUM}).\footnote{We cite the experiments with the strongest limits.  Other searches for these processes can be found in the Particle Data Group listings \cite{Amsler:2008zzb_PDG}.}  LFV in the first and third generations (``\lfvet '') is also possible, although current experimental limits on \taueg\ (from the BaBar collaboration \cite{Aubert:2009tk_BABAR}), and \taueee\ (from the BELLE collaboration \cite{Miyazaki:2007zw_BELLE}) are several orders of magnitude weaker than their \lfvem\ counterparts.  The gap between the \lfvem\ and \lfvet\ limits will continue to widen with the next generation of experiments searching for \lfvem , such as Fermilab's proposed \textit{Mu2e} experiment \cite{Mu2e_proposal} and the MEG experiment at PSI \cite{Adam:2009ci}. 

In this study, we examine the prospects for \lfvet\ searches at a prospective electron-ion collider.  Given the current status of experimental limits just described, one may ask whether  incremental ($\sim$ two orders of magnitude) improvements in \lfvet\ sensitivities are still useful as a probe of new physics.  The answer is certainly yes if there exist models in which processes such as \taueg\ and \etau\ are enhanced by several orders of magnitude relative to \mueg\ and \emu\ (or \mue) conversion.  One example of such a model is discussed in \cite{Ellis:2002fe}, in which the authors use a particular parameterization of the minimal supersymmetric seesaw model to find regions of parameter space which suppress the \mueg\ branching fraction and enlarge the \taueg\ branching fraction.  Another model in which \lfvet\ can be enhanced is the $SU(5)$ GUT with leptoquarks introduced in \cite{Dorsner:2005fq} and further studied in \cite{FileviezPerez:2008dw}.  In this model, the leptoquark couplings are tied to the neutrino masses and mixing matrix, with the result that in the quasi-degenerate neutrino mass regime the cross section for \etau\ can be relatively large despite strong constraints on \mue\ (we will describe this analysis in a later section of the paper). In light of these scenarios, the theoretical motivation for our analysis is clearly present.

In what follows, we analyze the prospective physics reach of a search for \lfvet\ at the proposed electron-ion collider (EIC), a high luminosity accelerator facility with the primary goal of exploring several open questions in QCD, such as the properties of the gluon distribution in nucleons and high-density quark-gluon matter \cite{EIC_whitepaper}.  In the eponymous electron-ion collisions at the EIC, the \lfvet\ process of interest is \etau\ conversion.  The \etau\ process can occur in leptoquark models via tree-level interactions, and so stronger signals may be expected relative to other models of \lfvet\ that induce lepton flavor violation through loop effects (e.g., doubly-charged Higgs loops in the left-right symmetric model).  Therefore, in our initial analysis of \lfvet\ at the EIC presented in this paper, we consider searches for leptoquark-induced \etau\ events.  For completeness, we adopt the general Buchmuller-Ruckl-Wyler leptoquark parameterization, described in \secref{sec:lqmodel}.  

As we will discuss in \secref{sec:crosssections}, the ZEUS and H1 experiments at HERA searched for \etau\ leptoquark events and placed upper limits on the ratio of the leptoquark-quark-lepton couplings divided by the squared leptoquark mass.  In addition to the direct searches for leptoquarks at HERA, experimental limits on \taueg\ and other rare decays (e.g., $\tau\rightarrow \pi e$,\ \taueee, and decays of $K$ and $B$ mesons) also place constraints on the same couplings-over-mass ratios \cite{Davidson:1993qk}.  At the time the HERA results were published, the limits from \taueg\ were weaker than limits from other rare processes and were not relevant for the analyses by the ZEUS and H1 collaborations.  In 2010, a stronger limit on the \taueg\ decay was published by the BaBar collaboration \cite{Aubert:2009tk_BABAR}, and so in \secref{sec:tauegamma}, we update the leptoquark limits using this most recent experimental result.  

Given an EIC operating at 90 \gev\ center-of-mass energy and with 10 \ifb\ of integrated luminosity, we will show in \secref{sec:results} that the EIC could set limits on the leptoquark coupling-over-mass ratios that surpass the current best limits from the HERA experiments by as much as nearly two orders of magnitude.  We will also show that the EIC could compete with or surpass the updated leptoquark limits from \taueg\ for a subset of the quark flavor-diagonal leptoquark couplings. Given this potential physics reach, it is interesting to ask whether there exist theoretically well-motivated scenarios in which \etau\ conversion could be observed at the EIC despite the much stronger limits on LFV from \mueg\ and \mue\ . To that end, in \secref{sec:neutrinos}, we discuss the interesting connection between one of the BRW leptoquarks and the model presented in \cite{Dorsner:2005fq}.  The leptoquark of interest is unconstrained by \mueg\ and \taueg, and using the results of the analysis in \cite{FileviezPerez:2008dw} we show that this leptoquark can yield \etau\ cross sections within reach of the EIC and still be compatible with the limits on neutrino masses and mixing angles and \mue\ conversion.  In \secref{sec:concl}, we will summarize our results and comment on a few considerations relevant for undertaking an experimental search for \etau\ at the EIC.  We will briefly make the point that searches for leptoquarks in \etau\ events at the EIC are complementary to leptoquark searches at hadron colliders like the Tevatron and LHC.  

\section{Leptoquark Framework}\label{sec:lqmodel}

We use the Buchmuller-Ruckl-Wyler (BRW) leptoquark parameterization \cite{Buchmuller:1986zs} as the framework for our \lfvet\ analysis.  The BRW parameterization catalogs all possible renormalizable and $SU(3)_C\times SU(2)_L\times U(1)_Y$ invariant interactions between (scalar or vector) leptoquarks and SM fermions.  These interactions are given by the Lagrangian in \eqnref{eq:lqlagrang}.
\begin{equation}\label{eq:lqlagrang}
\begin{aligned}
&\mathcal{L}_{LQ} &\ =\ & \mathcal{L}_{F=0} + \mathcal{L}_{\abs{F}=2}\\
&\mathcal{L}_{F=0} &\ =\ & h_{1/2}^L\overline{u}_R\ell_L S_{1/2}^L + h_{1/2}^R\overline{q}_L\epsilon e_R S_{1/2}^R + \tilde{h}_{1/2}^L\overline{d}_R\ell_L \tilde{S}_{1/2}^L + h_0^L\overline{q}_L\gamma_\mu\ell_L {V_0^L}^\mu\\
&\ &\ &\  + h_0^R\overline{d}_R\gamma_\mu e_R V_0^{R\mu} + \tilde{h}_0^R\overline{u}_R\gamma_\mu e_R\tilde{V}_0^{R\mu} + h_1^L\overline{q}_L\gamma_\mu\vec{\tau}\ell_L\vec{V}_1^{L\mu} + \mathrm{h.c.}\\
&\mathcal{L}_{\abs{F}=2} &\ =\ & g_0^L\overline{q}_L^c\epsilon\ell_L S_0^L + g_0^R\overline{u}_R^c e_R S_0^R + \tilde{g}_0^R\overline{d}_R^c e_R \tilde{S}_0^R + g_1^L\overline{q}_L^c\epsilon\vec{\tau}\ell_L\vec{S}_1^L + g_{1/2}^L\overline{d}_R^c\gamma_\mu \ell_L V_{1/2}^{L\mu}\\
&\ &\ &\ +g_{1/2}^R\overline{q}_L^c\gamma_\mu e_R V_{1/2}^{R\mu} + \tilde{g}_{1/2}^L\overline{u}_R^c\gamma_\mu\ell_L\tilde{V}_{1/2}^{L\mu} + \mathrm{h.c.}
\end{aligned}
\end{equation}
In \eqnref{eq:lqlagrang}, $q_L$ and $\ell_L$ are the $SU(2)$ doublet quarks and leptons, $u_R,\ d_R,\ e_R$ are the $SU(2)$ singlet quarks and charged lepton, $\epsilon$ is the $SU(2)$ antisymmetric tensor ($\epsilon_{12} = -\epsilon_{21} = +1$), $\vec{\tau} = \of{\tau_1, \tau_2, \tau_3}$ are the Pauli matrices, and the charge conjugated fermion is defined as $\psi^c \equiv C\overline{\psi}^T = i\gamma_2\gamma_0\overline{\psi}^T$ in the Dirac basis for the $\gamma$ matrices.  Color, $SU(2)$, and flavor (generation) indices have been suppressed.  The leptoquarks are characterized by their fermion number, their spin, the chirality of their coupling to leptons, and their gauge group quantum numbers.  The leptoquarks carry fermion number $F=3B+L$ equal to 0 or $\pm 2$.  We follow the notation used in the recent literature where spin-0 leptoquarks are $S$ and spin-1 are $V$, the subscript indicates the $SU(2)$ quantum number (0 for a singlet, 1/2 for a doublet, 1 for a triplet), the superscript $L,R$ indicates the chirality of the lepton coupling to the leptoquark, and a tilde ($\tilde{\ }$) is used to distinguish between leptoquarks which have different hypercharges but are otherwise identical.  The dimensionless coupling constants $g$ and $h$ (which we assume to be real) carry the same lepton chirality and $SU(2)$ labels as their associated leptoquarks.  Lepton flavor violation can arise if the couplings --- which are matrices in flavor space --- have non-zero off-diagonal elements.

We will also require the interactions between the BRW leptoquarks and the photon.  The photon interactions arise from the Lagrangian kinetic terms with $SU(2)_L\times U(1)_Y$ covariant derivatives acting on the leptoquark fields \cite{Belyaev:2005ew}:  
\begin{align}
&\mathcal{L}_{kinetic}^{(scalar)} = \of{D_{\mu}S}^\dagger\of{D^\mu S}\label{eq:scalarkineticterm}\ \ ,\\
&\mathcal{L}_{kinetic}^{(vector)} = -\frac{1}{2}\of{D_{\mu}V_{\nu} - D_{\nu}V_{\mu}}^\dagger\of{D^\mu V^\nu - D^\nu V^\mu}\label{eq:vectorkineticterm}\ \ .
\end{align}
The covariant derivative is given by 
\begin{equation}\label{eq:covariantderiv}
D_{\mu} = \partial_\mu + ig\vec{T}\cdot \vec{W}_\mu + ig'\frac{Y}{2}B_\mu\ \ ,
\end{equation}
where the $T^a$ are the generator matrices for the $SU(2)$ representation occupied by the leptoquarks (singlet, doublet\footnote{Note that the doublets must be in the $\overline{2}$ representation given the form of the Lagrangian in \eqnref{eq:lqlagrang}.  E.g., explicitly writing the $SU(2)$ indices, $\overline{u}_R {\ell_L}_i {S^L_{1/2}}_i$ shows that the $i=2$ component of the leptoquark multiplet couples to the electron and must have the opposite $T^3$ eigenvalue to be $SU(2)$ invariant.}, or triplet).  The photon interaction for a scalar leptoquark is given by
\begin{equation}\label{eq:lqscalarphoton}
\mathcal{L}_{LQ,\gamma}^{(scalar)} = ieQ_{LQ}\lb\of{\partial_\mu S^\dagger}S - S^\dagger\of{\partial_\mu S}\rb A^\mu\ \ ,
\end{equation}
where $Q_{LQ}$ is the electric charge of the leptoquark.  

For the vector leptoquarks, interactions with the photon depend on the nature of these massive vector particles, i.e., whether or not the leptoquarks are gauge bosons of some beyond-the-SM symmetry group.  In addition to the interaction arising from \eqnref{eq:vectorkineticterm}, there can exist an anomalous magnetic moment coupling of the leptoquark to the photon, so the full interaction Lagrangian is
\begin{equation}\label{eq:lqvectorphoton}
\mathcal{L}_{LQ,\gamma}^{(vector)} = -ieQ_{LQ}\lp\lb \mathcal{V}_{\mu\nu}^\dagger V^\nu - \mathcal{V}_{\mu\nu}{V^\nu}^\dagger\rb A^\mu - \of{1-\kappa}V_{\mu}^\dagger V_{\nu}F^{\mu\nu}\rp
\end{equation}
where the leptoquark field strength tensor $\mathcal{V}^{\mu\nu}$ is given by
\begin{equation}\label{eq:lqfieldtensor}
\mathcal{V}^{\mu\nu} \equiv \partial^\mu V^\nu - \partial^\nu V^\mu 
\end{equation}
and $F^{\mu\nu}$ is the usual photon field strength tensor.  If the leptoquarks are gauge bosons (as in the case of some $SU(5)$ GUTs, e.g.), then $\kappa=0$ and the resulting photon interaction is a three-gauge-boson vertex, the result of spontaneous symmetry breaking of the higher gauge group containing both the leptoquarks and the photons to $U(1)_{EM}$.  (Also, if the leptoquarks are gauge bosons, \eqnref{eq:vectorkineticterm} is replaced by the appropriate kinetic term for the gauge bosons of the larger symmetry group.)  This question of the gauge nature of the vector leptoquarks will have further implications for our analysis, particularly in the calculation of the \taueg\ limits (see \secref{sec:tauegamma}).  Finally, the electric charges of the scalar and vector leptoquarks which appear in the photon interaction terms are easily determined from \eqnref{eq:lqlagrang} (also, see Table 1 in \cite{Belyaev:2005ew}).

\section{Cross Section Calculations for \etau}\label{sec:crosssections}

Electron to tau conversion in an $e^-p$ deep inelastic scattering process is the \lfvet\ signal at the EIC which we consider in our analysis.  In the BRW leptoquark parameterization, such a process occurs via tree level partonic interactions.  In $e^-p$ collisions, $F=0$ type leptoquarks couple to antiquarks in the $s$-channel and quarks in the $u$-channel, while $\abs{F}=2$ type leptoquarks couple to quarks in the $s$-channel and antiquarks in the $u$-channel (see \figref{fig:etaufeyndiagrams}).  If the leptoquark mass is much larger than the center of mass energy, $M_{LQ} \gg \sqrt{s}$, the momentum dependence of the leptoquark propagator can be neglected, effectively shrinking the partonic interaction to a four-fermion vertex.  The cross section then depends only on the ratio of the leptoquark couplings divided by the leptoquark mass.  The total inclusive cross section for $e^-+p\rightarrow\tau^-+X$ with a single intermediate leptoquark is given (in the limit of massless quarks and leptons) by \cite{Chekanov:2002xz_ZEUS_old}
\begin{equation}\label{eq:etaucxn}
\begin{aligned}
&\sigma_{F=0} &\ =\ & \sum_{\alpha,\beta}\frac{s}{32\pi}\lb\frac{\lambda_{1\alpha}\lambda_{3\beta}}{M_{LQ}^2}\rb^2\left\{ \int dxdy\ x\overline{q}_\alpha\of{x,xs}f\of{y} 
	+ \int dxdy\ xq_\beta\of{x,-u}g\of{y}\right\}\ \ ,\\
&\sigma_{\abs{F}=2} &\ =\ & \sum_{\alpha,\beta}\frac{s}{32\pi}\lb\frac{\lambda_{1\alpha}\lambda_{3\beta}}{M_{LQ}^2}\rb^2\left\{ \int dxdy\ 
	xq_\alpha\of{x,xs}f\of{y}
	 + \int dxdy\ x\overline{q}_\beta\of{x,-u}g\of{y}\right\}\ \ .
\end{aligned}
\end{equation}
The functions $f$ and $g$ are defined in \eqnref{eq:fandg}.
\begin{equation}\label{eq:fandg}
f\of{y} = \left\{\begin{array}{cc} 1/2 & \mathrm{(scalar)}\\2\of{1-y}^2 & \mathrm{(vector)}\end{array}\right. \ \ , \ \ 
g\of{y} = \left\{\begin{array}{cc} \of{1-y}^2/2 & \mathrm{(scalar)}\\2 & \mathrm{(vector)}\end{array}\right.
\end{equation}
The parton distribution functions for the quarks and antiquarks are $q\of{x,Q^2}$ and $\overline{q}\of{x,Q^2}$, respectively, evaluated at momentum fraction $x$ and energy scale $Q^2$.  Also, $u = xs\of{y-1}$ and both $x$ and $y$ are integrated from 0 to 1.  The leptoquark couplings $\lambda_{1\alpha}$ and $\lambda_{3\beta}$ are the couplings $g$ and $h$ which appear in the Lagrangian of \eqnref{eq:lqlagrang} (additional factors of $-1$ and/or $\sqrt{2}$ may multiply these couplings, depending on the leptoquark $SU(2)$ representation --- see, e.g., Table 2 of \cite{Buchmuller:1986zs} and Table 1 of \cite{Belyaev:2005ew}).  The subscripts on the couplings $\lambda$ are generation indices: 1 and 3 for the electron and tau, and $\alpha$ and $\beta$ for the quarks/antiquarks.\footnote{Note that $\alpha$ is not always the initial state quark/antiquark; see \figref{fig:etaufeyndiagrams}.}  We refer to ratios with $\alpha=\beta$ as ``quark flavor-diagonal'' and those with $\alpha\neq\beta$ as ``quark flavor-off-diagonal''.  The ZEUS and H1 collaborations placed upper limits (at 95\% confidence level) on the ratio \lqratio\ for each type of BRW leptoquark and for all combinations of $\alpha$ and $\beta$ except in cases where the top quark was the only third-generation quark coupled to the leptoquark \cite{Chekanov:2002xz_ZEUS_old,Chekanov:2005au_ZEUS_new,Adloff:1999tp_H1_old,Aktas:2007ji_H1_new}.  To obtain these limits, several assumptions were made: only one type of leptoquark dominated the cross section, the leptoquark coupled only to left- or right-handed leptons but not both\footnote{This assumption was already made in writing the Lagrangian in \eqnref{eq:lqlagrang}.  Leptoquarks with identical quantum numbers, e.g. $S_0^L$ and $S_0^R$, could have identical couplings to left- and right-handed leptons: $g_0^L=g_0^R$.  In the original BRW parameterization \cite{Buchmuller:1986zs}, leptoquarks coupling to both left- and right-handed leptons were not differentiated.}, and leptoquarks in $SU(2)$ multiplets are degenerate in mass.  We make these assumptions in our analysis as well.

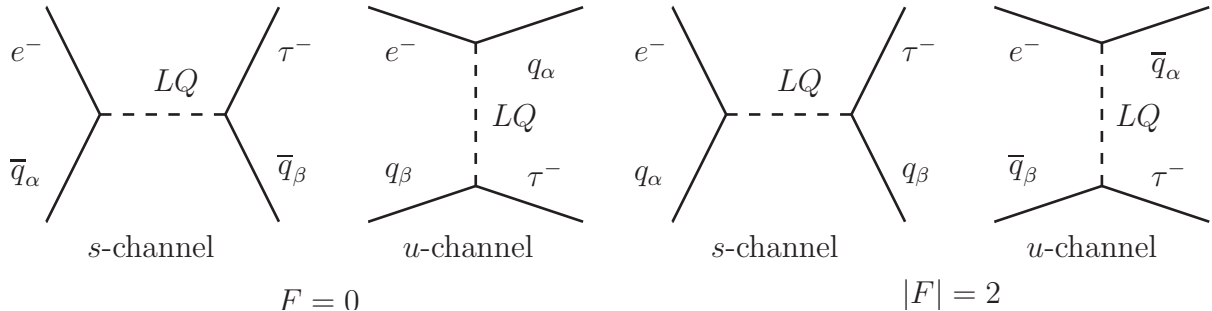
\begin{figure}
\begin{center}
\fcolorbox{white}{white}{
  \begin{picture}(451,113) (21,-39)
    \SetWidth{1.0}
    \SetColor{Black}
    \Line(33.709,72.895)(53.934,32.445)
    \Line(101.126,32.445)(121.351,72.895)
    \Line(101.126,32.445)(121.351,-8.006)
    \Line(33.709,-8.006)(53.934,32.445)
    \Line[dash,dashsize=4.214](53.934,32.445)(101.126,32.445)
    \Text(20.225,52.67)[lb]{\normalsize{\Black{$e^-$}}}
    \Text(20.225,5.478)[lb]{\normalsize{\Black{$\overline{q}_\alpha$}}}
    \Text(74.159,39.186)[lb]{\normalsize{\Black{$LQ$}}}
    \Text(121.351,52.67)[lb]{\normalsize{\Black{$\tau^-$}}}
    \Text(121.351,5.478)[lb]{\normalsize{\Black{$\overline{q}_\beta$}}}
    \Line(155.06,72.895)(195.511,59.412)
    \Line(155.06,-8.006)(195.511,5.478)
    \Line[dash,dashsize=4.214](195.511,59.412)(195.511,5.478)
    \Line(195.511,59.412)(235.961,72.895)
    \Line(195.511,5.478)(235.961,-8.006)
    \Text(161.802,52.67)[lb]{\normalsize{\Black{$e^-$}}}
    \Text(215.736,5.478)[lb]{\normalsize{\Black{$\tau^-$}}}
    \Text(161.802,5.478)[lb]{\normalsize{\Black{$q_\beta$}}}
    \Text(215.736,45.928)[lb]{\normalsize{\Black{$q_\alpha$}}}
    \Text(202.252,25.703)[lb]{\normalsize{\Black{$LQ$}}}
    \Line(269.67,72.895)(289.895,32.445)
    \Line(337.087,32.445)(357.312,72.895)
    \Line(337.087,32.445)(357.312,-8.006)
    \Line(269.67,-8.006)(289.895,32.445)
    \Line[dash,dashsize=4.214](289.895,32.445)(337.087,32.445)
    \Text(256.186,52.67)[lb]{\normalsize{\Black{$e^-$}}}
    \Text(256.186,5.478)[lb]{\normalsize{\Black{$q_\alpha$}}}
    \Text(310.12,39.186)[lb]{\normalsize{\Black{$LQ$}}}
    \Text(357.312,52.67)[lb]{\normalsize{\Black{$\tau^-$}}}
    \Text(357.312,5.478)[lb]{\normalsize{\Black{$q_\beta$}}}
    \Line(391.021,72.895)(431.472,59.412)
    \Line(391.021,-8.006)(431.472,5.478)
    \Line[dash,dashsize=4.214](431.472,59.412)(431.472,5.478)
    \Line(431.472,59.412)(471.922,72.895)
    \Line(431.472,5.478)(471.922,-8.006)
    \Text(397.763,52.67)[lb]{\normalsize{\Black{$e^-$}}}
    \Text(451.697,5.478)[lb]{\normalsize{\Black{$\tau^-$}}}
    \Text(397.763,5.478)[lb]{\normalsize{\Black{$\overline{q}_\beta$}}}
    \Text(451.697,45.928)[lb]{\normalsize{\Black{$\overline{q}_\alpha$}}}
    \Text(438.213,25.703)[lb]{\normalsize{\Black{$LQ$}}}
    \Text(121.351,-41.715)[lb]{\normalsize{\Black{$F=0$}}}
    \Text(357.312,-41.715)[lb]{\normalsize{\Black{$\abs{F}=2$}}}
    \Text(49.192,-21.489)[lb]{\normalsize{\Black{$s$-channel}}}
    \Text(168.544,-21.489)[lb]{\normalsize{\Black{$u$-channel}}}
    \Text(285.153,-21.489)[lb]{\normalsize{\Black{$s$-channel}}}
    \Text(404.505,-21.489)[lb]{\normalsize{\Black{$u$-channel}}}
  \end{picture}
}
\end{center}
	\caption{Feynman diagrams for \etau\ scattering processes via leptoquarks which depend on the parameter \lqratio.  The partonic cross section is convoluted with the pdf of the initial state (anti)quark of each diagram.  See \eqnref{eq:etaucxn}.}
	\label{fig:etaufeyndiagrams}
\end{figure}

To determine the sensitivity of an  EIC  search for \lfvet\ in \etau\ processes, we calculate an upper bound on the cross sections for the various leptoquarks using \eqnref{eq:etaucxn} and the most stringent limits on \lqratio\ from the ZEUS or H1 collaborations (or those rare process limits cited by the ZEUS and H1 collaborations).  We use the MSTW 2008 NLO set for the quark and antiquark proton p.d.f.s.\footnote{http://projects.hepforge.org/mstwpdf/}  From \eqnref{eq:etaucxn}, there is a one-to-one correspondence between the partonic sub-process cross section and the leptoquark ratio \lqratio.  Given a number for the sub-process cross section, we calculate the leptoquark ratio and then scale (i.e., divide) the leptoquark ratio by the HERA/rare process limit.  We define this scaled leptoquark ratio as the variable $z$.  Thus, for a given cross section there is a unique value of $z$.  In other words, $z$ is the fractional reduction in the leptoquark ratio relative to the HERA/rare process limit.  Results of these calculations will be presented in \secref{sec:results} after we discuss limits from \taueg.

\section{Updating \taueg~Limits}\label{sec:tauegamma}

As mentioned in the introduction, the BaBar collaboration has published a stronger limit \cite{Aubert:2009tk_BABAR} on the branching fraction for \taueg\ since the time of the ZEUS and H1 analyses (which did not make use of the weaker contemporaneous \taueg\ limit).  The experimental bound on \taueg\ implies additional constraints on the ratios \lqratio\ independent of the limits from HERA and other rare processes such as $\tau\rightarrow 3e,\ \tau\rightarrow \pi e$, and $B$ and $K$ decays.  However, the \taueg\ bound only applies to those leptoquark ratios where $\alpha=\beta$ (``quark flavor-diagonal'') since the \taueg\ process proceeds via a loop with a single quark and a leptoquark, shown in \figref{fig:tauegammadiagrams}.  In this section, we will use the recent limit from BaBar to calculate new leptoquark limits from \taueg.

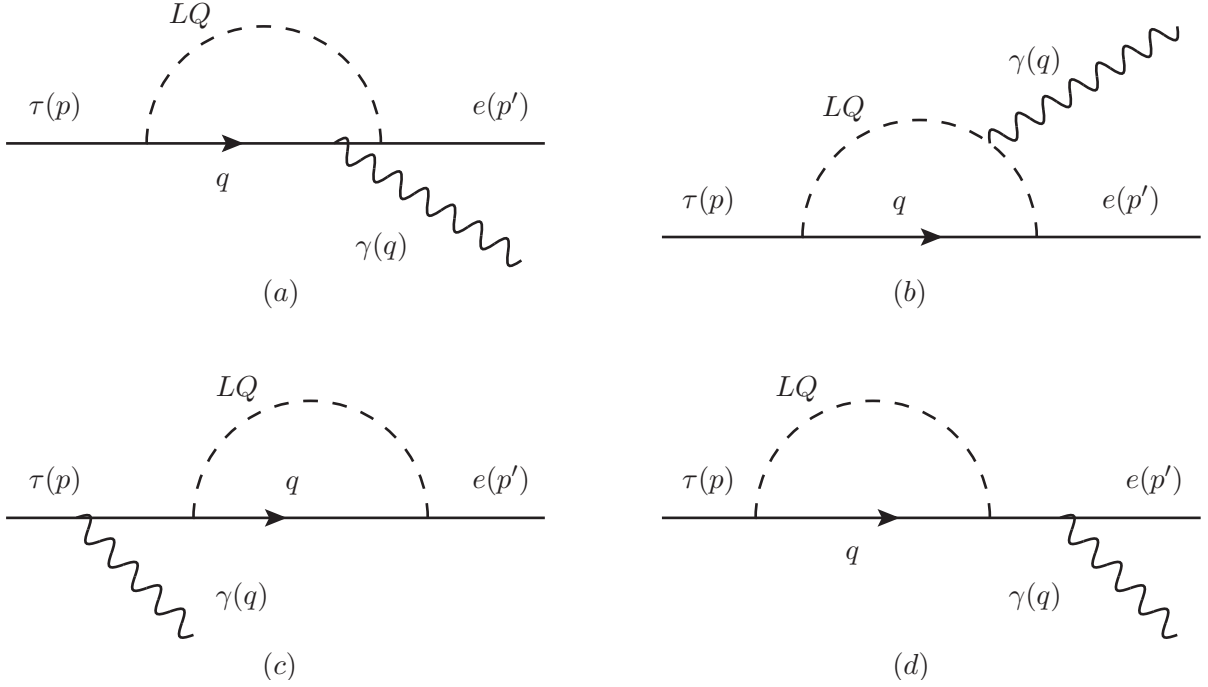
\begin{figure}
\begin{center}
\fcolorbox{white}{white}{
  \begin{picture}(451,255) (43,-32)
    \SetWidth{1.0}
    \SetColor{Black}
    \Line[arrow,arrowpos=0.42,arrowlength=6,arrowwidth=2.4,arrowinset=0.2](44.135,167.711)(247.153,167.711)
    \Arc[dash,dashsize=5.517,clock](141.231,167.711)(44.135,-180,-360)
    \Photon(167.711,167.711)(238.326,123.577){4.138}{7}
    \Text(52.961,176.538)[lb]{{\Black{$\tau(p)$}}}
    \Text(105.923,211.846)[lb]{{\Black{$LQ$}}}
    \Text(123.577,150.057)[lb]{{\Black{$q$}}}
    \Text(176.538,123.577)[lb]{{\Black{$\gamma(q)$}}}
    \Text(220.673,176.538)[lb]{{\Black{$e(p')$}}}
    \Line[arrow,arrowpos=0.5,arrowlength=6,arrowwidth=2.4,arrowinset=0.2](44.135,26.481)(247.153,26.481)
    \Arc[dash,dashsize=5.517,clock](158.884,26.481)(44.135,-180,-360)
    \Photon(70.615,26.481)(114.75,-17.654){4.138}{5}
    \Text(52.961,35.308)[lb]{{\Black{$\tau(p)$}}}
    \Text(123.577,70.615)[lb]{{\Black{$LQ$}}}
    \Text(150.057,35.308)[lb]{{\Black{$q$}}}
    \Text(123.577,-8.827)[lb]{{\Black{$\gamma(q)$}}}
    \Text(220.673,35.308)[lb]{{\Black{$e(p')$}}}
    \Line[arrow,arrowpos=0.42,arrowlength=6,arrowwidth=2.4,arrowinset=0.2](291.288,26.481)(494.307,26.481)
    \Arc[dash,dashsize=5.517,clock](370.73,26.481)(44.135,-180,-360)
    \Photon(441.345,26.481)(485.48,-17.654){4.138}{5}
    \Text(300.115,35.308)[lb]{{\Black{$\tau(p)$}}}
    \Text(335.422,70.615)[lb]{{\Black{$LQ$}}}
    \Text(361.903,8.827)[lb]{{\Black{$q$}}}
    \Text(423.691,-8.827)[lb]{{\Black{$\gamma(q)$}}}
    \Text(467.826,35.308)[lb]{{\Black{$e(p')$}}}
    \Line[arrow,arrowpos=0.5,arrowlength=6,arrowwidth=2.4,arrowinset=0.2](291.288,132.404)(494.307,132.404)
    \Arc[dash,dashsize=5.517,clock](388.384,132.404)(44.135,-180,-360)
    \Photon(414.865,167.711)(485.48,211.846){4.138}{7}
    \Text(300.115,141.23)[lb]{{\Black{$\tau(p)$}}}
    \Text(353.076,176.538)[lb]{{\Black{$LQ$}}}
    \Text(379.557,141.23)[lb]{{\Black{$q$}}}
    \Text(423.691,194.192)[lb]{{\Black{$\gamma(q)$}}}
    \Text(458.999,141.23)[lb]{{\Black{$e(p')$}}}
    \Text(141.23,105.923)[lb]{{\Black{$(a)$}}}
    \Text(379.557,105.923)[lb]{{\Black{$(b)$}}}
    \Text(379.557,-35.308)[lb]{{\Black{$(d)$}}}
    \Text(141.23,-35.308)[lb]{{\Black{$(c)$}}}
  \end{picture}
}
\end{center}
	\caption{Leptoquark loops contribute to \taueg\ decay.}
	\label{fig:tauegammadiagrams}
\end{figure}

In general, the amplitude for a $\tau\rightarrow e\gamma^*$ process can be written \cite{Hisano:1995cp,RamseyMusolf:2006vr}
\begin{equation}\label{eq:tauegammaamp}
\begin{aligned}
\mathcal{M}_{\tau\rightarrow e\gamma^*} =& e {\epsilon^*}^\nu \overline{u}_e\of{p'}\lb\of{q^2\gamma_\nu - q_\nu\slashed{q}}\of{A_1^L P_L + A_1^R P_R}\right.\\
&\qquad\qquad\qquad\quad \left. + im_\tau q^\alpha\sigma_{\nu\alpha}\of{A_2^L P_L + A_2^R P_R}\rb u_\tau\of{p}\ \ .
\end{aligned}
\end{equation}
For a real photon ($q^2=0$), only the magnetic moment term containing the coefficients $A_2^L$ and $A_2^R$ will contribute to $\abs{\mathcal{M}}^2$; the branching ratio for the \taueg\ process is then 
\begin{equation}\label{eq:brtauegamma}
Br\of{\tau\rightarrow e\gamma} \equiv \frac{\Gamma\of{\tau^-\rightarrow e^-\gamma}}{\Gamma\of{\tau^-\rightarrow e^-\overline{\nu}_e\nu_\tau}}
	= \frac{48\pi^3\alpha_{EM}}{G_\mu^2} \of{\abs{A_2^L}^2 + \abs{A_2^R}^2}\ \ .
\end{equation}
In \eqnref{eq:brtauegamma}, $G_\mu$ is the Fermi constant obtained from muon decay.

We calculate the contributions to $A_2^L$ and $A_2^R$ for each scalar leptoquark (vector leptoquarks will be discussed later) by computing the amplitude for the diagrams in \figref{fig:tauegammadiagrams} according to the Feynman rules obtained from the leptoquark Lagrangian (\eqnref{eq:lqlagrang}) and from the photon interactions in \eqnref{eq:lqscalarphoton}.  In the limit of zero electron mass, only the first two diagrams of \figref{fig:tauegammadiagrams} are needed; however, all four diagrams are necessary to cancel the loop divergences and thus avoid introducing a tree level counterterm for a $\tau,e,\gamma$ vertex.  To evaluate the loop integrals, we approximate $m_\tau^2/M_{LQ}^2\simeq 0$ and expand in powers of $m_q^2/M_{LQ}^2$.  With this procedure, our results are valid for all quarks running in the loop, including top quarks.  For all of the scalar leptoquarks, the expressions for $A_2^L$ and $A_2^R$ have the same basic structure: the $A_2$ coefficient with the same left- or right-handed label as the leptoquark is zero, while the other coefficient is given by
\begin{equation}\label{eq:lqA2}
A_2^{\of{L,R}} = -\frac{1}{16\pi^2}\frac{N_c}{6}\sum_{\alpha=1}^3\of{\frac{\lambda_{1\alpha}\lambda_{3\alpha}}{M_{LQ}^2}}\lc\of{\mathcal{Q}_q 
	+ \frac{\mathcal{Q}_{LQ}}{2}} + \frac{m_q^2}{M_{LQ}^2}\of{\frac{\mathcal{Q}_q}{2}\lb 11+6\ln{\frac{m_q^2}{M_{LQ}^2}}\rb - \mathcal{Q}_{LQ}}\rc\ \ .
\end{equation}
In \eqnref{eq:lqA2}, $N_c = 3$ is the number of colors and $\mathcal{Q}_q$ ($\mathcal{Q}_{LQ}$) is the sum of the electric charges of all the quarks (leptoquarks) appearing in the loop --- note that leptoquarks in $SU(2)$ doublets or triplets can couple to both up- and down-type quarks, with different electric charges for each member of the multiplet.  The quark charge contribution is from diagram (a) of \figref{fig:tauegammadiagrams}, while the leptoquark charge contribution comes from diagram (b).  The leptoquark charges are determined by their $SU(2)$ eigenvalue and hypercharge, $Q=T_3+Y/2$, and can be inferred from the Lagrangian in \eqnref{eq:lqlagrang}.  Complete results for the individual scalar leptoquarks are given in \appref{sec:appendix}.  We note that our results for the scalar leptoquarks, \eqnref{eq:lqA2}, agree with the results in \cite{Davidson:1993qk} and \cite{Gabrielli:2000te}.\footnote{The authors of \cite{Davidson:1993qk} assumed massless quarks and therefore do not have a term proportional to $m_q^2/M_{LQ}^2$.  The author of \cite{Gabrielli:2000te} assumes that the couplings $\lambda$ are unitary, so the first term in \eqnref{eq:lqA2} which is independent of the quark mass vanishes when summing over the quark generations.}

The uncertain gauge nature of the vector leptoquarks presents a difficulty in calculating the \taueg\ loop amplitudes.  In particular, the interaction between the photon and leptoquark depends on the unknown parameter $\kappa$.  Moreover, if the vector leptoquarks are gauge bosons ($\kappa = 0$), their propagator can be written (in the 't Hooft--Feynman gauge) as 
\begin{equation}\label{eq:masslessvector}
\frac{-ig^{\mu\nu}}{k^2-M_{LQ}^2+i\vareps}\ \ ,
\end{equation}
but if the vector leptoquarks are not gauge bosons, then their propagator is
\begin{equation}\label{eq:massivevector}
\frac{-i}{k^2-M_{LQ}^2+i\vareps}\of{g^{\mu\nu}-\frac{k^\mu k^\nu}{M_{LQ}^2}}\ \ .
\end{equation}
The second term in \eqnref{eq:massivevector} introduces extra divergences in the loop graphs of \figref{fig:tauegammadiagrams} that do not cancel like the divergences of the scalar leptoquark diagrams.  The authors of \cite{Davidson:1993qk} argued that the second term in the propagator of \eqnref{eq:massivevector} can be neglected for the purpose of extracting upper limits on the leptoquark coupling-to-mass ratios: since the additional divergences introduced by this term go like powers of a large (model-dependent) cutoff scale, ignoring the extra contribution to the propagator results in conservative (i.e., weaker) upper limits. 

Here, we adopt a different perspective.  The manifest non-renormalizability of the massive vector leptoquark theory leads to several options. It is possible, for example, that the vector leptoquarks are gauge bosons of a larger gauge group and that they receive their masses through spontaneous breaking of the gauge symmetry down to that of the Standard Model. In this context, one would presumably need to include loop contributions involving other massive degrees of freedom associated with the extended gauge group, and it is not clear without specifying a model whether these contributions will add to, or cancel against, the leptoquark loops. Alternatively, one might consider the leptoquarks (not necessarily as gauge bosons) as part of a low-energy effective theory valid below a scale $\Lambda\sim M_{LQ}$. In this case, there would exist a set of higher dimension LFV operators with {\em a priori} unknown coefficients determined by physics above the scale $\Lambda$. In this situation, it is possible to derive \lq\lq naturalness" bounds on the leptoquark couplings by requiring that the individual contributions from cut-off dependent loop amplitudes and  higher dimension operators be no larger than the \taueg\ bounds. Depending on which option one chooses, one may derive different limits on the coupling-to-mass ratios. In short, it is not possible to derive theoretically robust constraints using the particle content of \eqnref{eq:lqlagrang} alone. Therefore, we do not attempt to infer  \taueg\ limits for vector leptoquarks and restrict our attention to the scalar leptoquarks.

Experiments \cite{Amsler:2008zzb_PDG,Aubert:2009tk_BABAR} have given
\begin{equation}\label{eq:taubranchingfracs}
\frac{\Gamma\of{\tau^-\rightarrow e^-\overline{\nu}_e\nu_\tau}}{\Gamma\of{\tau\ total}} = 0.1785\ \ ,\ \ \frac{\Gamma\of{\tau^-\rightarrow e^-\gamma}}{\Gamma\of{\tau\ total}} < 3.3\times 10^{-8}
\end{equation}
so we take $Br\of{\tau\rightarrow e\gamma} \leq 1.85\times 10^{-7}$.  From \eqnref{eq:brtauegamma} and \eqnref{eq:lqA2}, it is evident that this experimental upper bound on the branching ratio for the \taueg\ decay places an upper limit only on a sum over generations of the leptoquark ratios \lqratiod; furthermore, the terms proportional to the quark mass contain additional powers of the unknown leptoquark mass in the denominator.  So, we will first neglect the terms proportional to $m_q^2/M_{LQ}^2$ since we assume the leptoquarks have large masses, $M_{LQ}\gg\mathcal{O}\of{100~GeV}$.  The quark mass is expected to make a significant contribution to the $A_2$ coefficients only if the quark in question is a top quark, and we will comment on the effect of including the top quark mass at the end.  Then, to extract an upper limit on the ratio for a single generation, we consider two different options.  The first option is to assume that each individual quark generation saturates the \taueg\ limit.  This assumption gives a weaker (larger) upper limit on the ratio for each generation $\alpha$.  The second, more ``democratic'', option assumes that all generations contribute equally to the \taueg\ branching ratio, giving a stronger (smaller) upper limit on the leptoquark ratio for each generation.  In both options, the limit is equivalent for all generations.  These two options determine a range of values for the upper bound on the leptoquark ratios \lqratiod\ arising from the experimental \taueg\ limit.  \Tabref{tab:tauegammaexample} gives an example of these limits for the leptoquark $\tilde{S}_0^R$.  

Including the top quark mass will result in an increase in the calculated upper bound on \lqratiod; the increase is mostly independent of the leptoquark type (since all the leptoquarks coupling to the top quark have charges that are $\mathcal{O}\of{1}$) but dependent on the leptoquark mass itself and which of the two options is used to calculate the upper bound on \lqratiod.  We show in \tabref{tab:quarkmasseffect} the percent increases in the leptoquark ratio upper bounds under various assumptions.  By neglecting the top quark mass, we get stronger upper limits on the leptoquark ratios from \taueg; i.e., the ratios \lqratiod\ are smaller when the top quark mass is neglected.  This is preferable for making comparisons with the ability of the EIC to probe these leptoquark ratios since smaller limits from \taueg\ will force more conservative (larger) estimates of the integrated luminosity necessary for the EIC to surpass the \taueg\ limits.  In the next section, we combine the \taueg\ limits with the cross section calculations discussed in the previous section and present our results.

\begin{table}[t]
\centering
\begin{tabular}{|c|c|c|}\hline 
 & Option 1 & Option 2\\\hline 
\lqratiod & $0.481~{TeV}^{-2}$ & $0.160~{TeV}^{-2}$\\\hline 
$\alpha=1,\ z=$ & 1.2 & 0.40\\\hline 
$\alpha=2,\ z=$ & 0.074 & 0.025\\\hline 
$\alpha=3,\ z=$ & 0.032 & 0.011\\\hline 
\end{tabular}
\caption{\taueg\ upper limits on \lqratiod\ for the leptoquark $\tilde{S}_0^R$.  Option 1 assumes each generation individually saturates the \taueg\ limit, while Option 2 assumes all three generations have equal contributions.  We compute the scaled leptoquark ratio $z$ by dividing the \taueg\ limit on \lqratiod\ by the limit from HERA and other rare processes.  $z>1$ implies that the bound from HERA/rare processes is already stronger than the new limit from \taueg.  These limits appear as the vertical dashed lines in \figref{fig:samplecxnplot} and \figref{fig:cxnstauegamma}.}
\label{tab:tauegammaexample}
\end{table}

\begin{table}[t]
\centering
\begin{tabular}{|c|c|c|}\hline 
 & $M_{LQ} = 500 GeV$ & $M_{LQ} = 1500 GeV$\\\hline 
Option 1 & +25\% & +10\%\\\hline 
Option 2 & +7\% & +3\%\\\hline 
\end{tabular}
\caption{Estimates of the effect of including the top quark mass in calculating the upper limits on \lqratiod\ from \taueg.  Including the top quark mass weakens (i.e., increases by the percentage indicated) the upper bounds on the leptoquark ratios \lqratiod\ calculated from the \taueg\ experimental bound.  The top quark mass effect depends on the leptoquark mass and the details of how the leptoquark ratio \lqratiod\ is calculated (described in the text as Option 1, in which a single quark generation saturates the upper bound, and Option 2, in which all three quark generations contribute equally to the upper bound).}
\label{tab:quarkmasseffect}
\end{table}

\section{Results and Discussion}\label{sec:results}

We now present our numerical results and discuss their implications for the EIC.  In principle, with 10 \ifb\ of integrated luminosity, the EIC could probe previously unexplored regions of parameter space for leptoquark-induced \etau\ events with a cross section of 0.1 $fb$ or greater.  For the specific case of \etau\ events involving leptoquarks, we make use of \eqnref{eq:etaucxn} to plot the partonic sub-process cross sections for individual BRW leptoquarks as a function of $z$, the leptoquark ratio \lqratio\ scaled by the corresponding HERA limit.  A center-of-mass energy of $\sqrt{s}=90$~\gev\ (roughly corresponding to, e.g., a 10~\gev\ electron beam and 200~\gev\ proton beam) is assumed.  We separate the contributions to the total inclusive cross section $\sigma\of{e^- p\rightarrow \tau^- X}$ from all combinations of initial and final state quark/antiquark generations; the separate contributions are proportional to \lqratio\ where $\alpha$ and $\beta$ are the quark generation numbers, as discussed above.  A value of $z = 1$ corresponds to the maximum cross section allowed by the limits from HERA.\footnote{We use the most restrictive value of \lqratio\ from the most recent ZEUS \cite{Chekanov:2005au_ZEUS_new} and H1 \cite{Aktas:2007ji_H1_new} analyses and the rare process results cited therein.  In fact, for all \etau\ leptoquark processes, the ZEUS limits are stronger than the H1 results, although this is not the case for the rare process limits.  The rare process limits used in the HERA analyses did not make use of the relatively weak bound on \taueg\ which existed at the time.}  For scalar leptoquark quark flavor-diagonal cases where $\alpha=\beta$, the \taueg\ upper limits on \lqratiod\ derived according to the two options discussed in \secref{sec:tauegamma} are also applicable; these bounds are also scaled by the corresponding HERA limit to obtain values for the variable $z$.  

\begin{figure}
	\includegraphics[width=.96\textwidth]{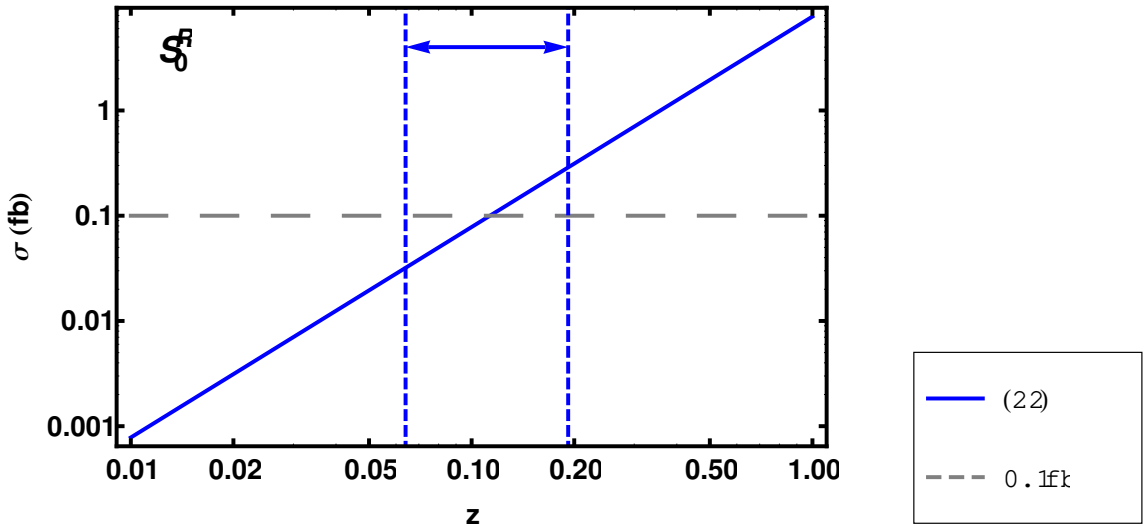}
	\caption{Cross section for the scalar leptoquark $S_0^R$ plotted according to \eqnref{eq:etaucxn} as a function of $z$, defined in the text as the leptoquark ratio \lqratio\ divided by the HERA/rare process limit on that ratio.  This plot shows the cross section with second generation initial and final state quarks, $\of{\alpha,\beta}=\of{2,2}$.  Upper bounds on $\lambda_{12}\lambda_{32}/M_{LQ}^2$ are calculated from the \taueg\ limit according to the two options discussed in the text; these upper bounds are indicated by the vertical dashed lines joined by the horizontal arrow.  The upper limit on $\lambda_{12}\lambda_{32}/M_{LQ}^2$ from \taueg\ may be as large as the vertical line on the right or as small as the vertical line on the left.}
	\label{fig:samplecxnplot}
\end{figure}

\begin{figure}
	\includegraphics[width=.48\textwidth]{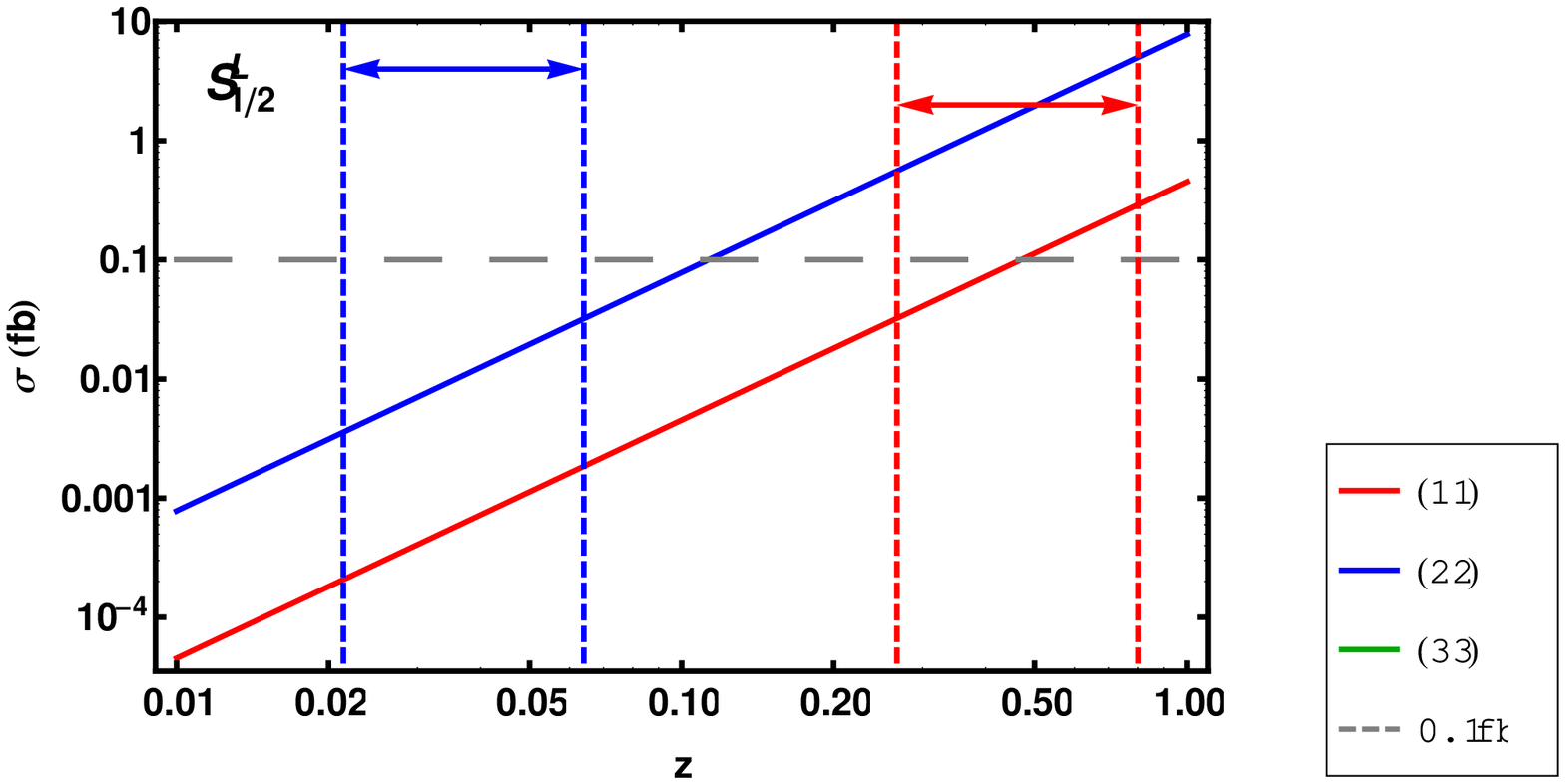}
	\includegraphics[width=.48\textwidth]{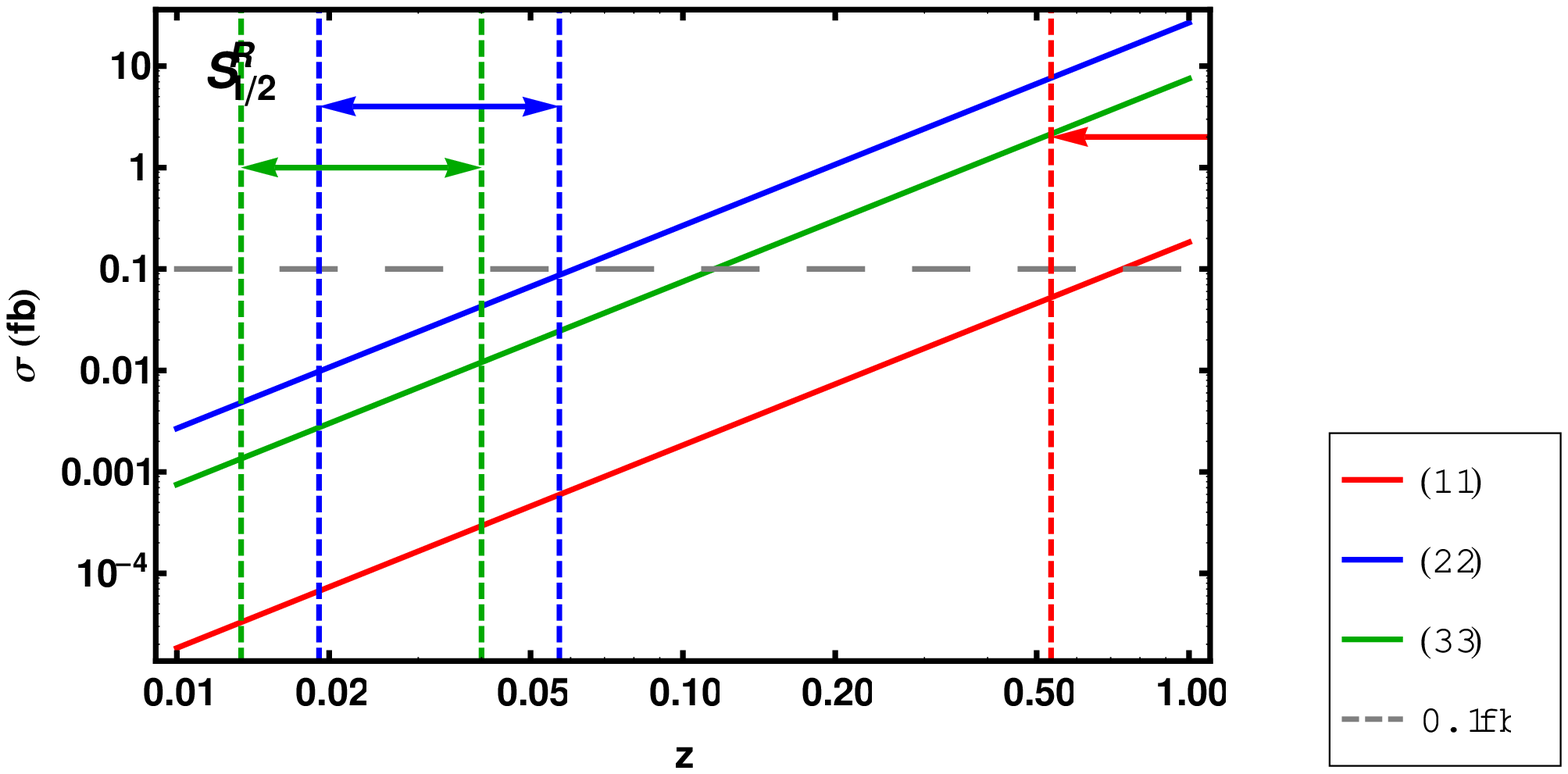}\\
	\includegraphics[width=.48\textwidth]{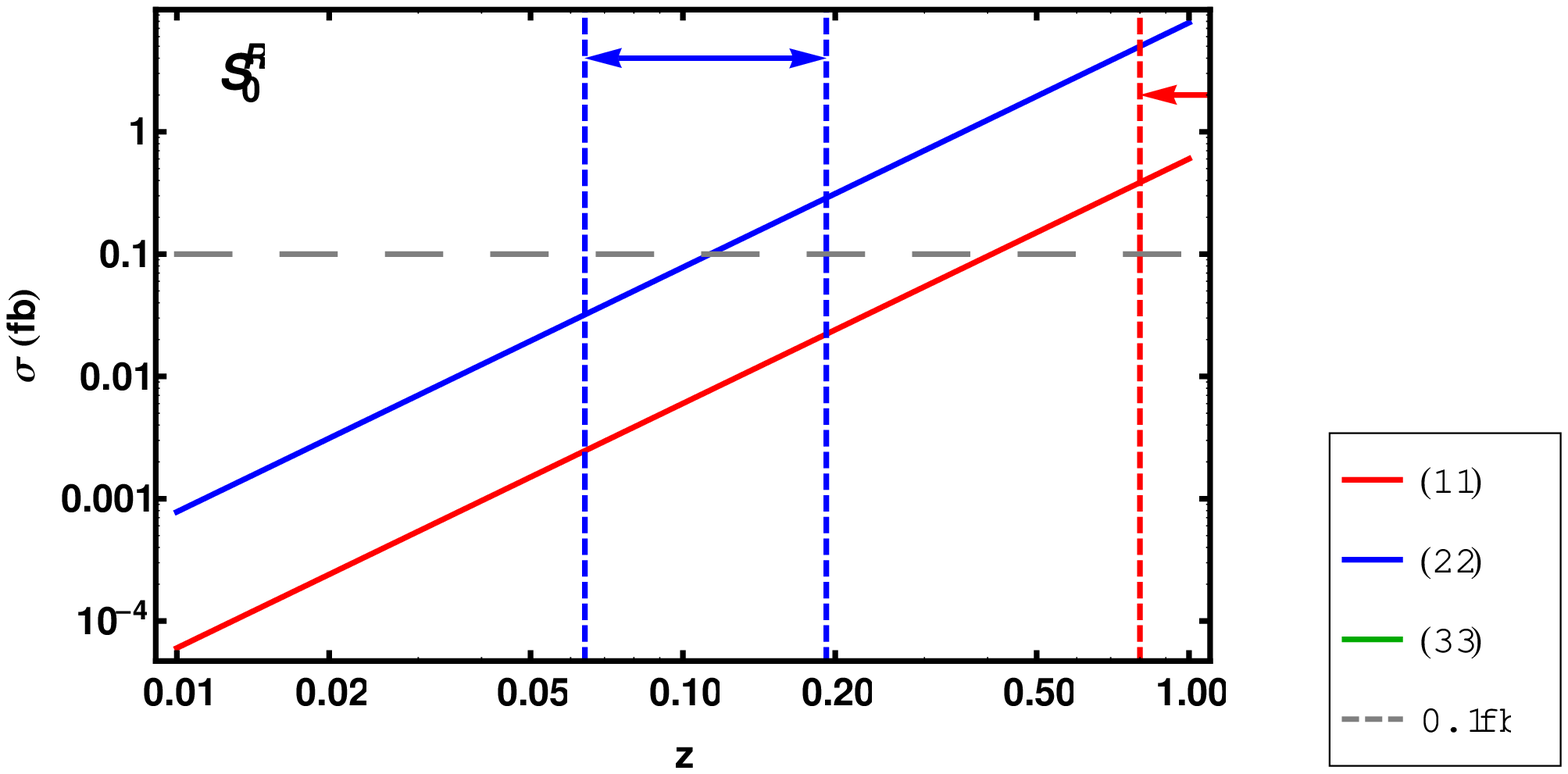}
	\includegraphics[width=.48\textwidth]{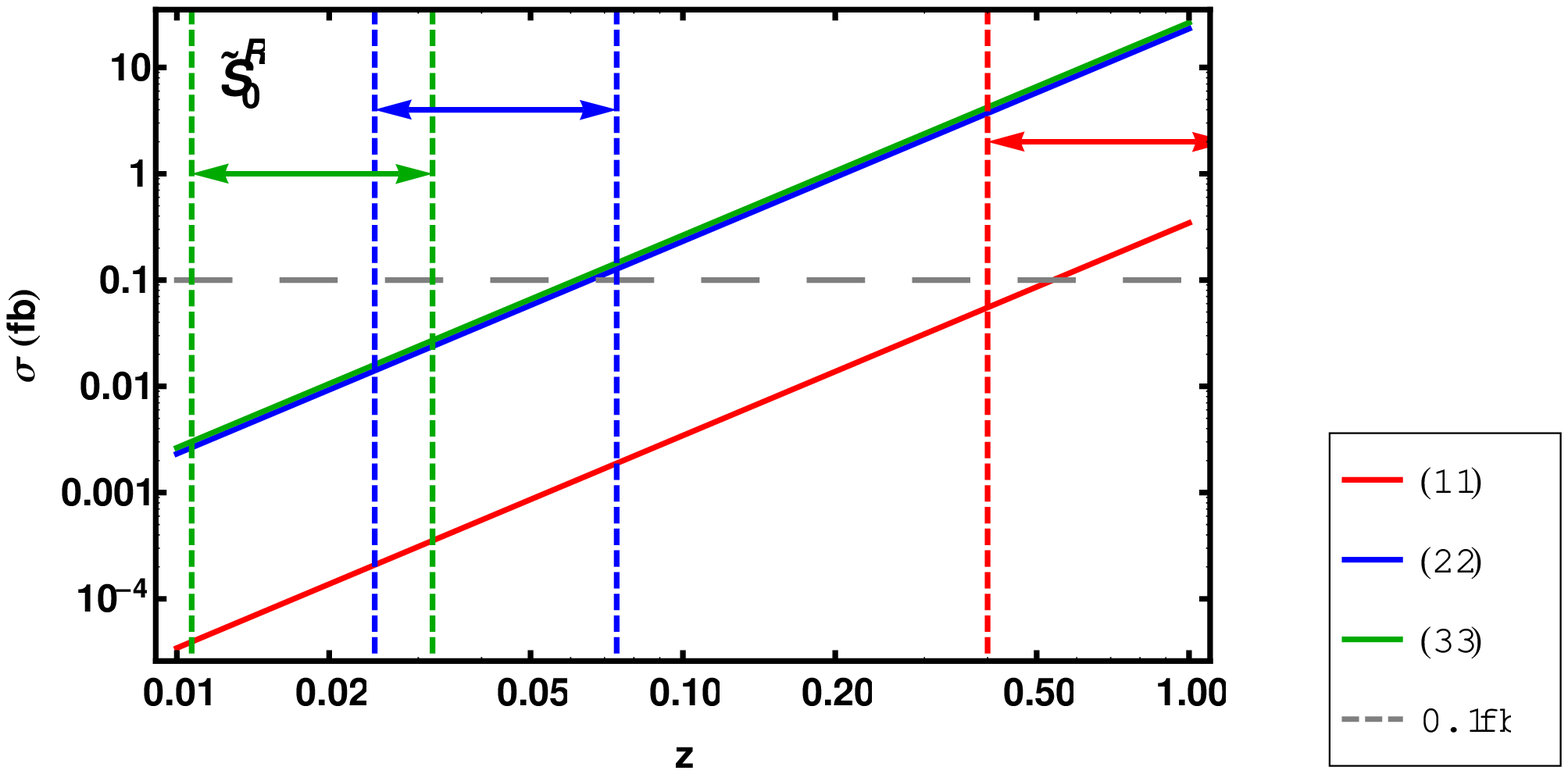}
	\caption{Cross sections for various scalar leptoquarks plotted according to \eqnref{eq:etaucxn} as a function of $z$, the scaled leptoquark ratio \lqratio.  These plots show the quark flavor-diagonal cross sections with $\of{\alpha,\beta} = \of{1,1},\of{2,2},\of{3,3}$.  The range of the leptoquark ratios \lqratiod\ that satisfy the \taueg\ limit are indicated by the vertical dashed lines and horizontal arrows.  Cross sections and \taueg\ limits for third generation quarks are not shown if the leptoquark couples exclusively to the top quark; no limit from the HERA analyses exists for these cases.}
	\label{fig:cxnstauegamma}
\end{figure}
\begin{figure}
	\includegraphics[width=.48\textwidth]{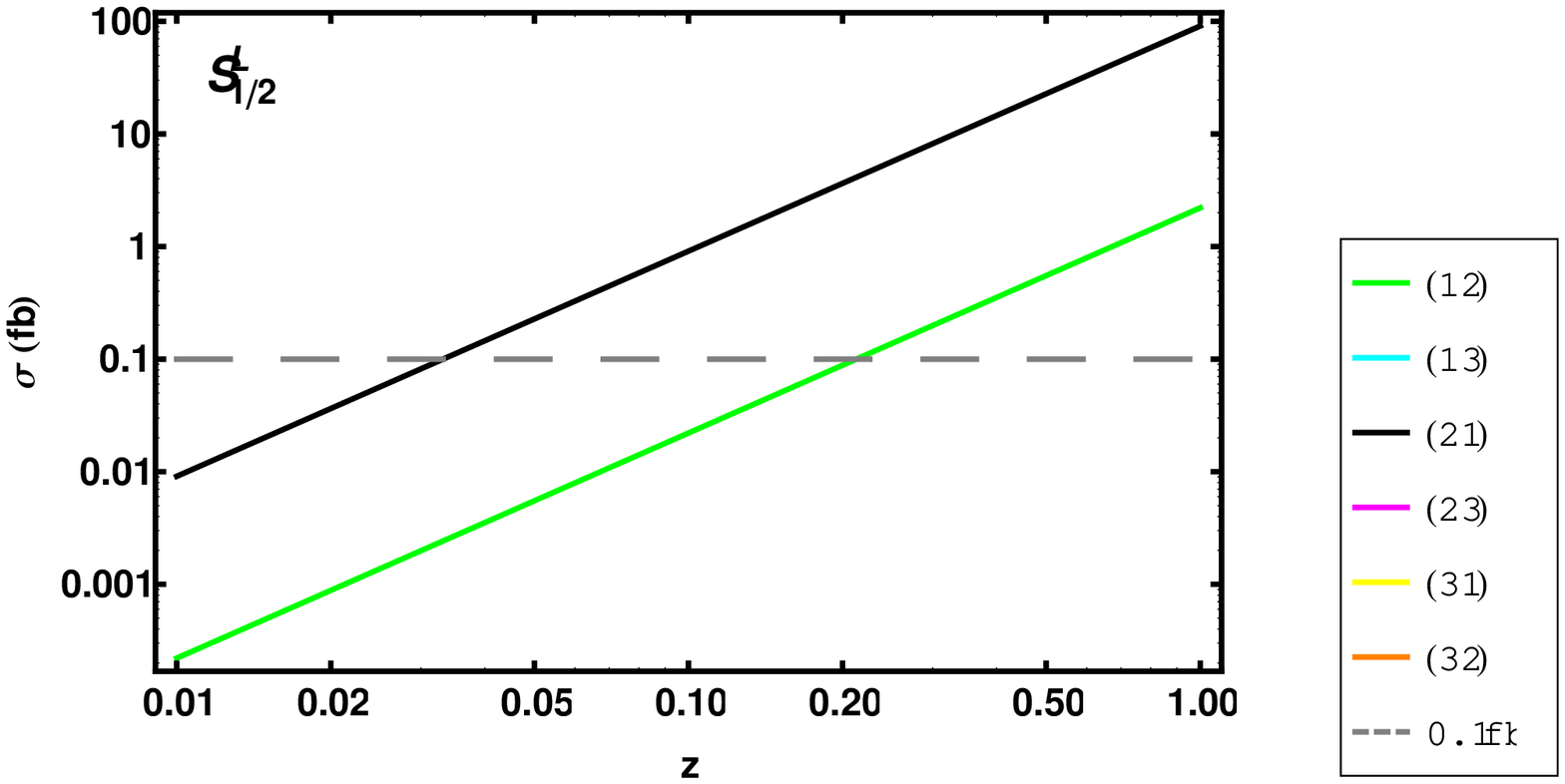}
	\includegraphics[width=.48\textwidth]{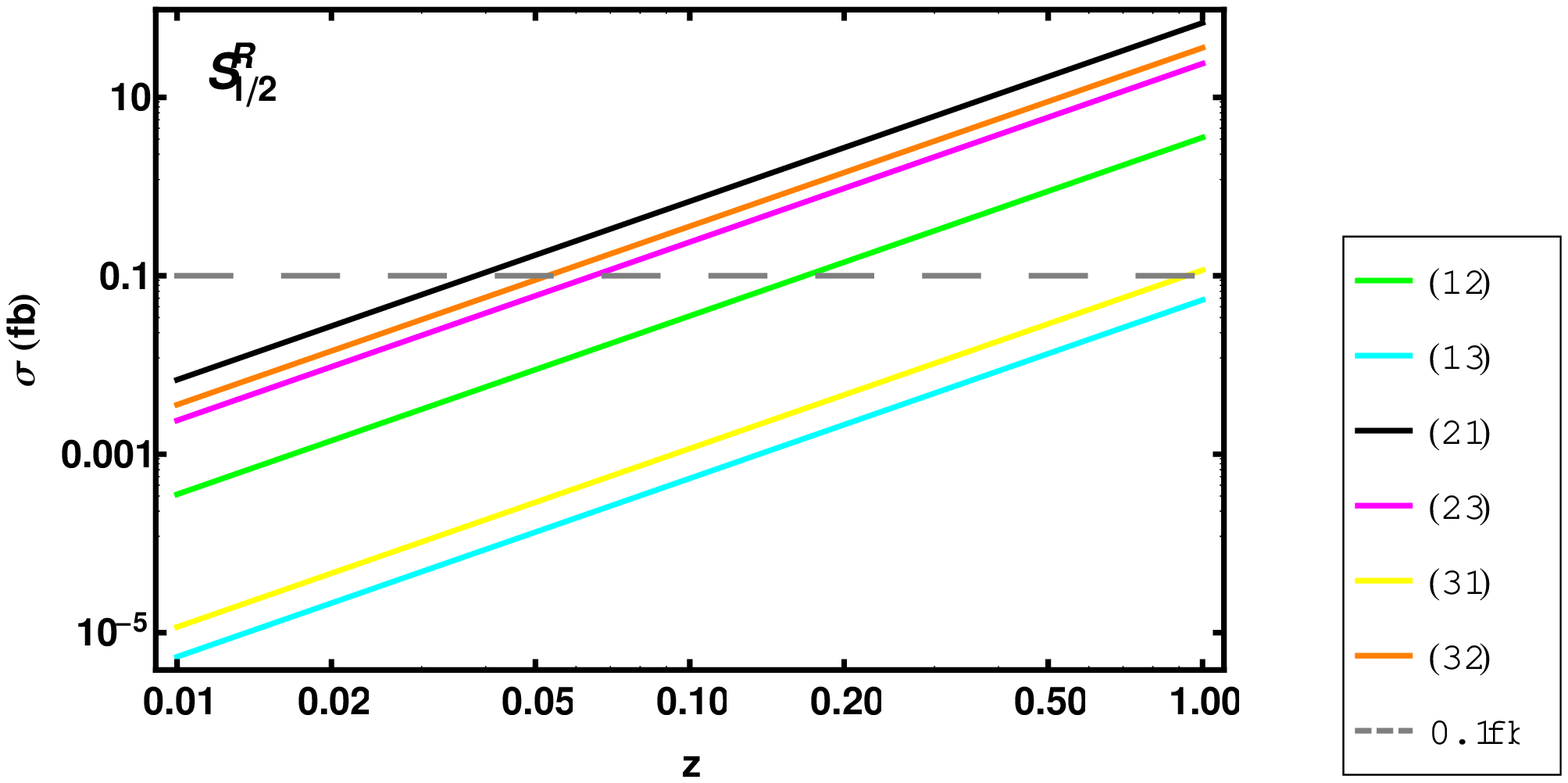}\\
	\includegraphics[width=.48\textwidth]{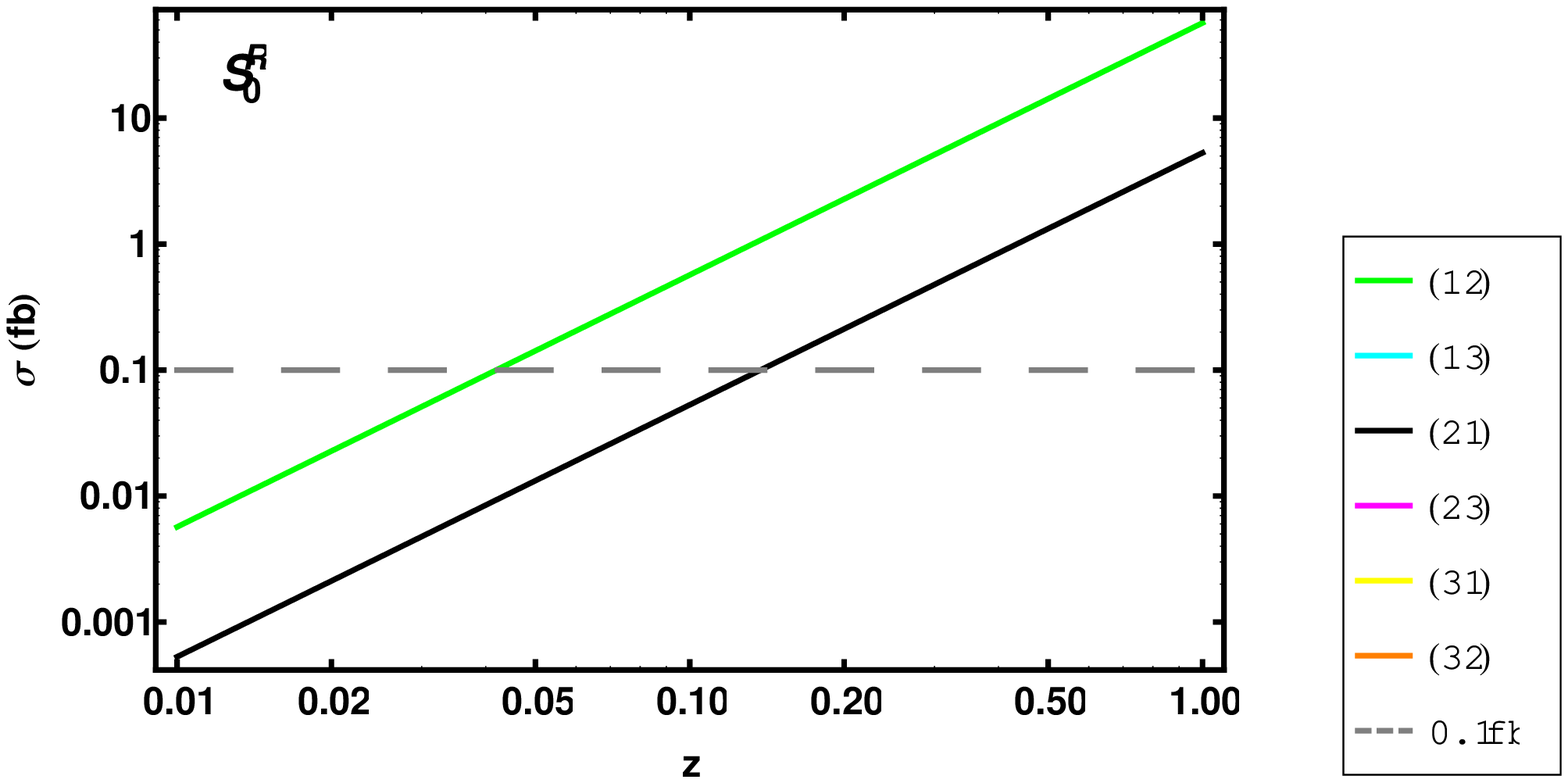}
	\includegraphics[width=.48\textwidth]{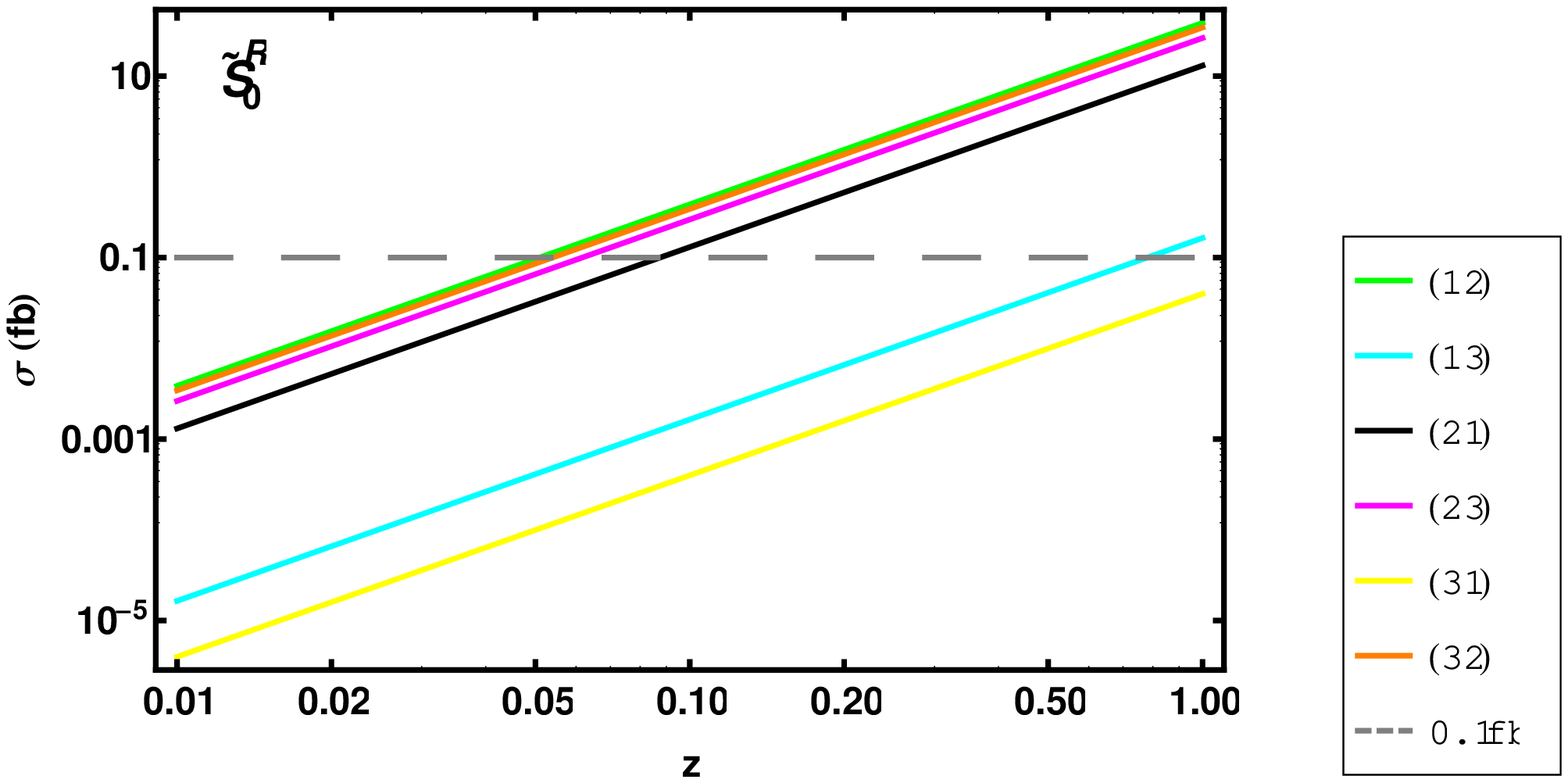}
	\caption{Cross sections for various scalar leptoquarks plotted according to \eqnref{eq:etaucxn} as a function of $z$.  These plots show the quark flavor-off-diagonal cross sections with $\alpha\neq\beta$.  Cross sections for third generation quarks are not shown if the leptoquark couples exclusively to the top quark; no limit from the HERA analyses exists for these cases.}
	\label{fig:cxnsnotauegamma}
\end{figure}

In \figref{fig:samplecxnplot}, we show for the particular leptoquark $S_0^R$ the cross section calculated according to \eqnref{eq:etaucxn} as a function of the leptoquark ratio \lqratio\ (scaled by the corresponding HERA limit to give the variable $z$).  We have chosen $\of{\alpha,\beta} = \of{2,2}$, corresponding to second generation quarks/antiquarks in the initial and final states (see \figref{fig:etaufeyndiagrams}).  Since $\alpha=\beta$ is ``quark flavor-diagonal'', limits from \taueg\ apply: the vertical dashed line on the right corresponds to the upper bound on $\lambda_{12}\lambda_{32}/M_{LQ}^2$ calculated according to Option 1 and the vertical dashed line on the left corresponds to Option 2, as discussed in \secref{sec:tauegamma}.  The upper bound on the \taueg\ branching ratio corresponds to an upper limit on $\lambda_{12}\lambda_{32}/M_{LQ}^2$ that may lie anywhere between the two vertical dashed lines (this region is indicated by the horizontal arrow).  The \taueg\ bounds on the leptoquark ratios have also been scaled by the HERA limit.  The horizontal dashed line indicates a 0.1~$fb$ cross section, which is roughly the level of sensitivity the EIC could achieve with 10~\ifb\ integrated luminosity.  If no \etau\ events were observed, the EIC could set a new limit on the ratio $\lambda_{12}\lambda_{32}/M_{LQ}^2$ given by the intersection of the diagonal line with the horizontal dashed line.  This occurs for $z\simeq 0.1$; hence, the new limit would be one order of magnitude smaller than the current limit.  The new limit on $\lambda_{12}\lambda_{32}/M_{LQ}^2$ would surpass the weaker \taueg\ upper bound from Option 1, but not the stronger limit of Option 2.

In \figref{fig:cxnstauegamma} and \figref{fig:cxnsnotauegamma}, we show the cross sections calculated for four different scalar leptoquarks, $S^L_{1/2}, S^R_{1/2}, S^R_0,$ and $\tilde{S}_0^R$, as a function of the scaled leptoquark ratio, $z$.  We have chosen to display results for these four leptoquarks because they demonstrate many general features which we will discuss below.  Each line shown in these plots corresponds to a contribution to the total inclusive cross section from a particular combination of quark generations $\of{\alpha,\beta}$, indicated in the legends.  In \figref{fig:cxnstauegamma}, we show the cross sections involving the quark flavor-diagonal $\alpha=\beta$ leptoquark ratios; the vertical dashed lines and horizontal arrows indicate the possible location of the \taueg\ upper bound, as discussed for \figref{fig:samplecxnplot}.  Plots for quark flavor-off-diagonal contributions to the inclusive cross section are shown in \figref{fig:cxnsnotauegamma}.

From these results for the cross sections, we can make several remarks which are generally true for most of the scalar leptoquarks.
\begin{enumerate}
\item It is clear that the present limits ($z=1$) on the ratios \lqratio\ involving first generation quarks are more stringent than those limits involving second and third generation quarks (these stronger limits are in many cases from the rare process limits other than \taueg\ rather than the direct searches at HERA); hence, the allowed cross sections for initial state first generation quarks are suppressed despite their larger p.d.f. contribution.  As a result, with 10 \ifb\ and $\sqrt{s}=90$~\gev, it is unlikely that the EIC would observe first generation quark \etau\ events if the leptoquark ratios are smaller than half their current limit (the cross section contributions from the $\of{\alpha,\beta}=\of{1,1}$ combination cross the 0.1~$fb$ sensitivity threshold near $z=0.5$).  To probe the first generation quark ratios to the level of one order of magnitude smaller, 100 to 1000~\ifb\ are likely necessary.  (However, the cross sections mixing the first and second generations in \figref{fig:cxnsnotauegamma} seem to be an exception to this observation.)  
\item For those leptoquark ratios \lqratio\ involving second and third generation quarks, the current limits are less stringent and so the EIC could potentially achieve an order of magnitude improvement, or more, with the given energy and luminosity; this is possible despite the suppression of these cross sections by the proton p.d.f.s.
\item We can now address how \etau\ searches at the EIC would fare in light of the most recent \taueg\ limit.  As \figref{fig:cxnstauegamma} shows, the present upper limit from HERA and/or other rare processes is roughly on the same order of magnitude as the limit on \lqratiod\ from \taueg\ for the first generation quarks (the \taueg\ upper bounds are near $z=1$).  A factor of two decrease in these limits attained by the EIC would achieve parity with the strongest \taueg\ upper bound from BaBar.  For the second and third generation quarks, the \taueg\ limits are presently much stronger than the HERA limits (where such limits exist --- recall that HERA did not set limits for leptoquarks coupling only to top quarks in the third generation, and the EIC would likewise be unable to probe such couplings).  Even the weaker of the two \taueg\ upper bounds we calculated seem to be generally out of reach of the EIC given our choice of integrated luminosity and center-of-mass energy. We emphasize, nonetheless, that these statements apply only to the quark flavor-diagonal combinations of scalar leptoquark couplings. Compared to present bounds, the EIC sensitivity to quark flavor off-diagonal combinations could substantially exceed that of other searches performed to date.
\item The analysis presented here is predicated on a choice of 90~\gev\ center-of-mass energy and 10~\ifb\ integrated luminosity.  Larger integrated luminosities will allow the EIC to be sensitive to smaller cross sections, and larger center-of-mass energies will increase all of the cross sections uniformly since $\sigma\propto s$ as in \eqnref{eq:etaucxn}.  The results of our analysis above are broadly improved by such changes.  For example, we observe from \figref{fig:cxnstauegamma} that 1000~\ifb\ of integrated luminosity would allow the EIC to improve all of the leptoquark limits by an order of magnitude or more --- even those limits involving the first generation quarks, as mentioned above --- and surpass the stronger \taueg\ limits for all three generations.
\end{enumerate}

Although the vector leptoquarks are problematic when calculating the \taueg\ limits, the vector leptoquark contributions to \etau\ are more straightforward.  The second term of the vector propagator in \eqnref{eq:massivevector} could be present in the amplitude for the \etau\ scattering diagrams in \figref{fig:etaufeyndiagrams}; however, the contribution to the cross sections from this term will be suppressed by additional powers of the large leptoquark mass and may safely be neglected.  Making use of \eqnref{eq:etaucxn}, we show in \figref{fig:vectorqd} and \figref{fig:vectorqod} the quark flavor-diagonal and quark flavor-off-diagonal cross sections, respectively, as a function of $z$ for two of the vector leptoquarks, $V_1^L$ and $\tilde{V}_{1/2}^L$.  Many of the same features discussed above for the scalar leptoquarks also generally hold true for the vector leptoquarks (with the principal exception that there is no straightforward application of the \taueg\ bound to the vector leptoquarks).  For example, as for the scalar leptoquarks, it is evident that the EIC could obtain an order of magnitude improvement on the limits for the second generation quark flavor-diagonal ratios \lqratiod\ with $\sqrt{s}=90$~\gev\ and 10~\ifb\ integrated luminosity.  Finally, we point out that the leptoquark $V_1^L$ displays a behavior not seen for the other leptoquarks presented so far: the cross sections involving couplings to first and second generation quarks, $\of{\alpha,\beta}=\of{1,2},\of{2,1}$, are highly suppressed.\footnote{The source of this suppression is the stringent limit on the decay $K\rightarrow\pi\nu\nu$.}  We conclude that the potential of the EIC to probe leptoquark-induced \lfvet\ is mostly independent of whether the leptoquarks are scalar or vector particles.

\begin{figure}
	\includegraphics[width=.48\textwidth]{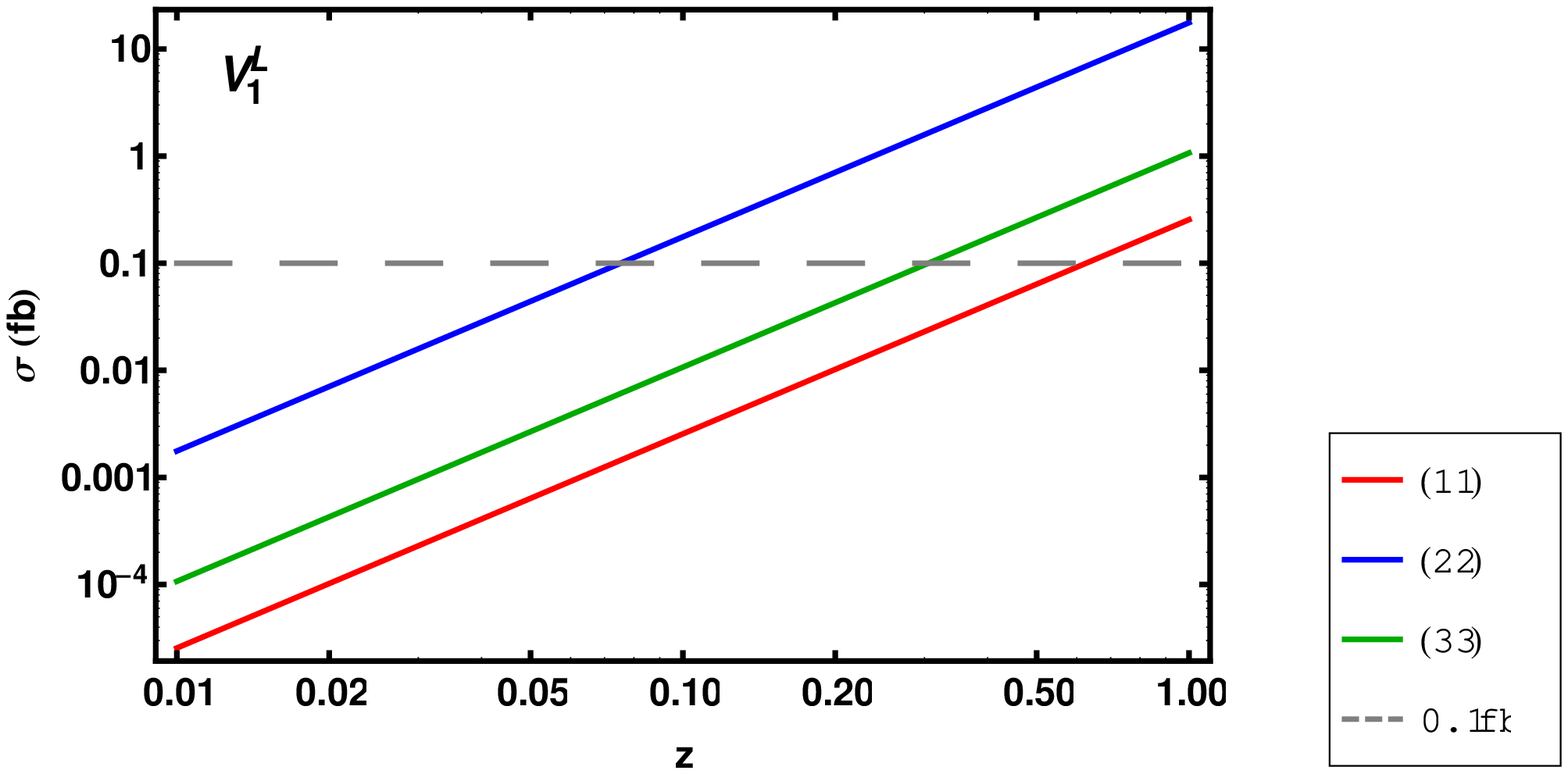}
	\includegraphics[width=.48\textwidth]{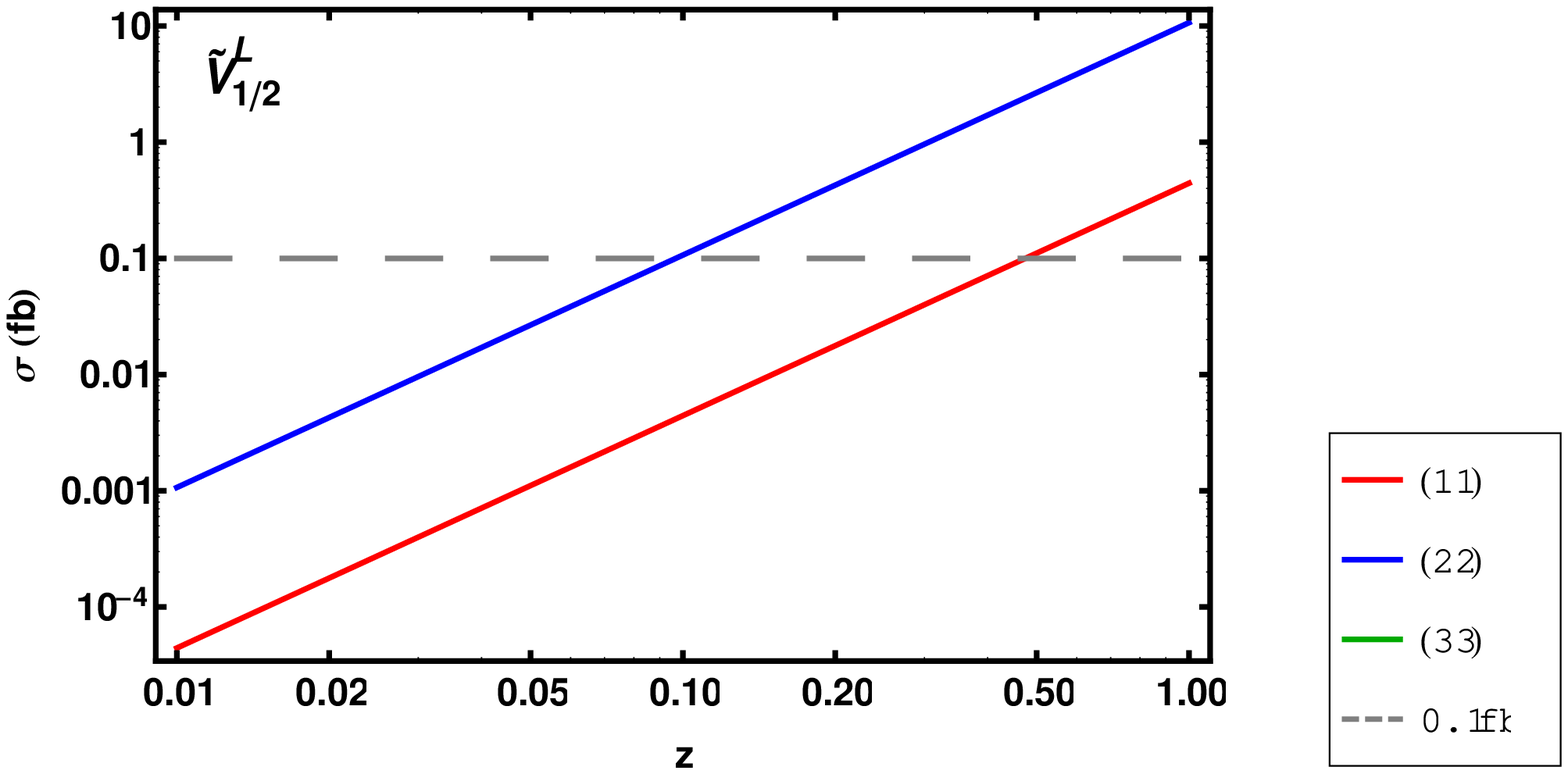}
	\caption{Analogous to \figref{fig:cxnstauegamma}, these plots show the quark flavor-diagonal cross sections for two vector leptoquarks.  We do not show limits from \taueg\ as discussed in the text.}
	\label{fig:vectorqd}
\end{figure}
\begin{figure}
	\includegraphics[width=.48\textwidth]{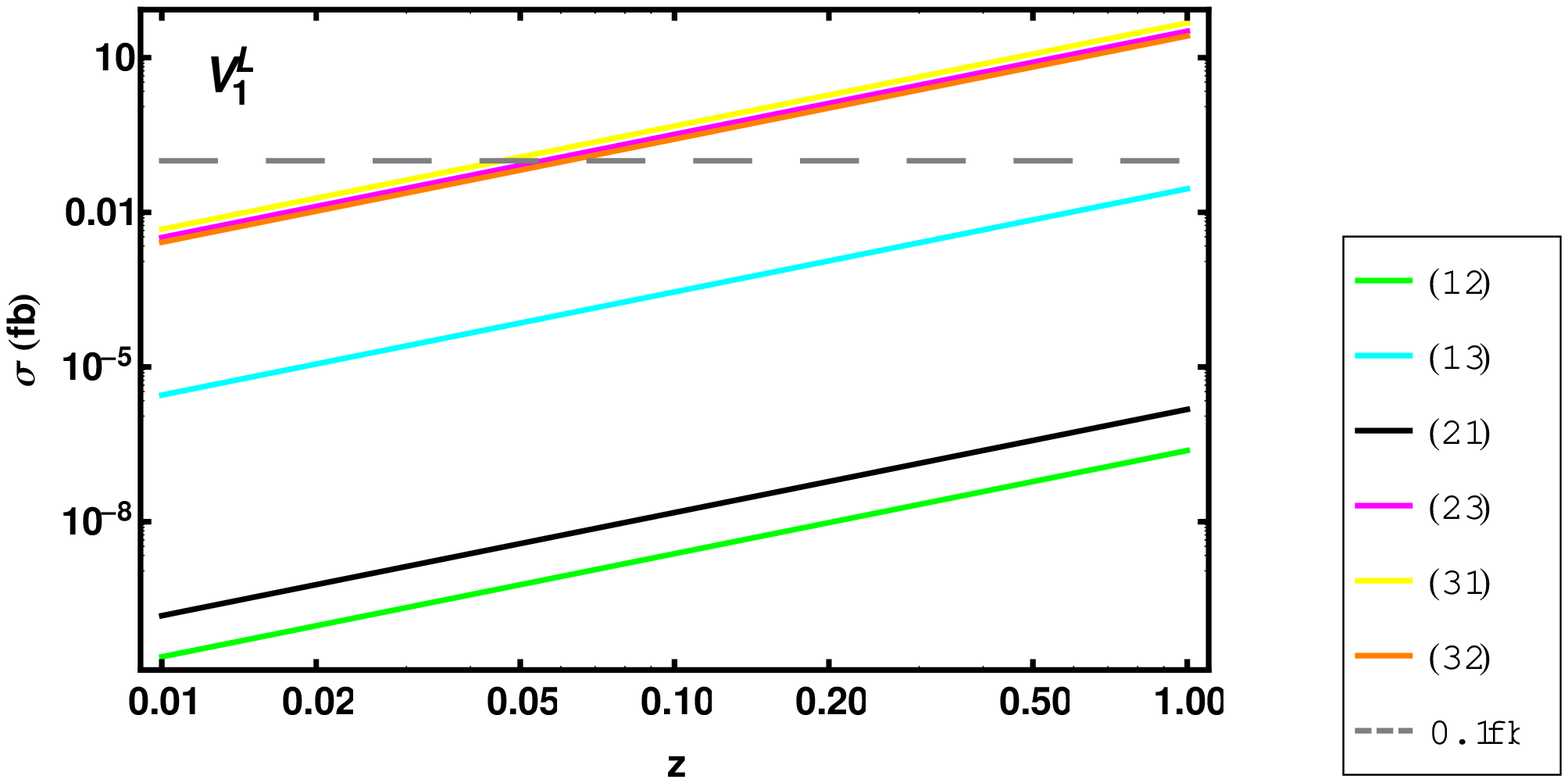}
	\includegraphics[width=.48\textwidth]{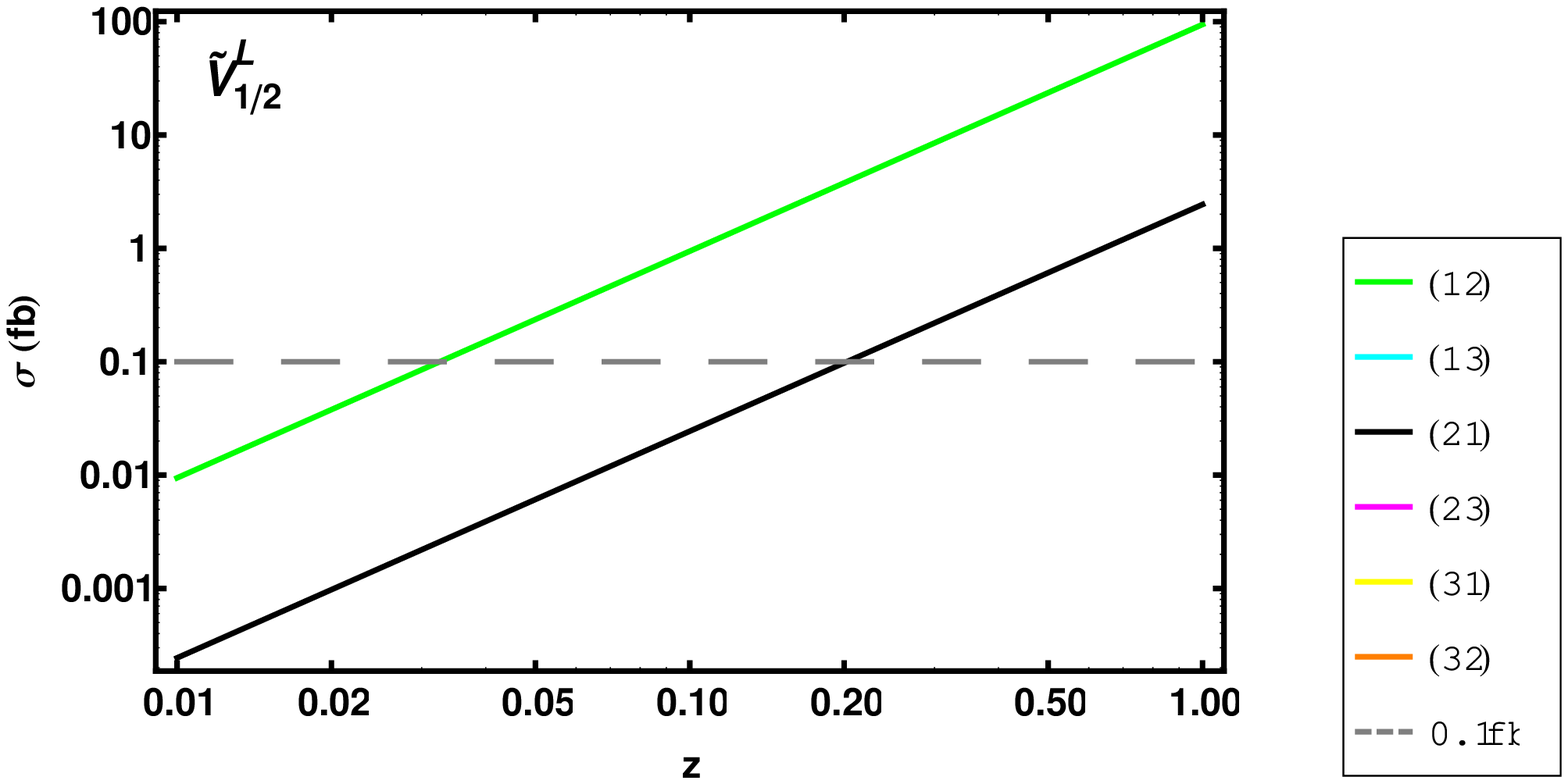}
	\caption{Analogous to \figref{fig:cxnsnotauegamma}, these plots show the quark flavor-off-diagonal cross sections for two vector leptoquarks.}
	\label{fig:vectorqod}
\end{figure}

\section{A Model Illustration: $\tilde{S}_{1/2}^L$ and Neutrinos}\label{sec:neutrinos}

One of the BRW leptoquarks is of particular interest because of its connection to the work in \cite{FileviezPerez:2008dw}.  The leptoquark studied by those authors ($\Phi_b$ in their notation) has the same spin and gauge group quantum numbers as the leptoquark \stilde.  The authors of \cite{FileviezPerez:2008dw} consider an $SU(5)$ GUT model in which this leptoquark arises; as a result of the $SU(5)$ symmetry, the coupling of this leptoquark to quarks and leptons can be constrained by the neutrino sector: by a particular choice of rotation matrices (in particular to avoid proton decay) and neglecting phases, the neutrino mass matrix determines the leptoquark couplings.  Given the experimental constraints on the squared neutrino mass differences and mixing angles, the authors perform a scan to determine allowed values of the leptoquark couplings as a function of the lightest neutrino mass in both normal and inverted neutrino mass hierarchies.  The authors introduce a dimensionful coupling parameter $Y_1^{ij} = \Gamma_1^{ij}\times v_\Delta$ where $\Gamma_1$ is the leptoquark coupling, equivalent to our $\lambda$, and $v_\Delta$ is the vacuum expectation value of the scalar triplet field $\Delta$ which implements the type-II seesaw mechanism.  The superscripts $i$ and $j$ refer to the quark and lepton generations, respectively (this is a reversal of our notation, $\lambda_{1\alpha}$ and $\lambda_{3\beta}$, where the first index refers to the lepton and the second index refers to the quark).  Their results are shown in figures 1, 2, and 3 of \cite{FileviezPerez:2008dw}.  The quasi-degenerate neutrino mass region, where the lightest neutrino mass is $\mathcal{O}\of{10^{-1}~eV}$, is of particular interest because of the several orders of magnitude separation in some of the leptoquark couplings $Y_1$.  In the quasi-degenerate region, satisfying the neutrino mass and mixing constraints forces the couplings $Y_1^{31},Y_1^{22},$ and $Y_1^{13}$ to be $\mathcal{O}\of{10^{-1}~eV}$, while $Y_1^{12}$ and $Y_1^{23}$ are $\mathcal{O}\of{10^{-3}~eV}$ and the remaining couplings are relatively unconstrained and can take on values between $10^{-3}$ and $10^{-1}~eV$.

Are there other constraints on this model's leptoquark couplings in addition to those imposed by the neutrino masses?  As \eqnref{eq:allLQAs} shows, this leptoquark's coupling-over-mass ratios are not constrained by the limit from \taueg\ (more precisely, the leading contribution to the \taueg\ branching fraction from \stilde\ is suppressed by $m_q^2/M_{LQ}^4$ rather than $1/M_{LQ}^2$, so the upper bounds on \lqratiod\ from \taueg\ are very weak).  Limits from \mueg\ are also evaded for the same reason.  We instead consider the \stilde\ leptoquark's contribution to \mue\ conversion in nuclei and derive constraints from this tightly constrained \lfvem\ process.  The \mue\ conversion rate is calculated from a model-independent effective Lagrangian in \cite{Kitano:2002mt} and can be written in terms of the leptoquark dimensionful couplings $Y_1$ as \cite{FileviezPerez:2008dw}
\begin{equation}\label{eq:mueconvrate}
\omega_{conv} = \frac{\abs{Y_1^{11}}^2\abs{Y_1^{12}}^2m_\mu^5}{4M_{LQ}^4v_\Delta^4}\of{V^{\of{p}}+2V^{\of{n}}}^2\ \ .
\end{equation}
The quantities $V^{\of{p}}$ and $V^{\of{n}}$ are overlap integrals of the muon and electron wave functions with the nucleon densities, calculated for various nuclei in \cite{Kitano:2002mt}.  Using these overlap integrals and muon capture rate for gold nuclei \cite{Kitano:2002mt} and the SINDRUM II collaboration's result for the ratio of the muon capture to conversion rates in gold nuclei \cite{Bertl:2006up_SINDRUM},
\begin{equation}\label{eq:muoncapture}
\begin{aligned}
&R_{\mu\rightarrow e}^{Au} = \frac{\Gamma\lb \mu^- + A\of{Z,N}\rightarrow e^- + A\of{Z,N}\rb}{\Gamma\lb \mu^- + A\of{Z,N}\rightarrow \nu + A\of{Z-1,N+1}\rb} = \frac{\omega_{conv}}{\omega_{capt}} < 7.0\times 10^{-13}\ \ \of{90\%\ \text{C.L.}}\ \ ,\\
&\omega_{capt} = 13.07\times 10^6 s^{-1}\ \ ,
\end{aligned}
\end{equation}
we can calculate limits on the leptoquark couplings $Y_1$.  In the quasi-degenerate region, $Y_1^{12}$ is already constrained to be $~10^{-3}~eV$ as mentioned above; hence, \eqnref{eq:mueconvrate} and \eqnref{eq:muoncapture} dictate that $Y_1^{11}\apprle 2\times 10^{-3}~eV$, two orders of magnitude smaller than the maximum allowed by the parameter scan in \cite{FileviezPerez:2008dw}.

Using the limits on the leptoquark couplings from both the neutrino sector and the \mue\ bound, we can compute the \etau\ cross section from \eqnref{eq:etaucxn}, replacing \lqratio\ with $Y_1^{\alpha 1}Y_1^{\beta 3}/\of{M_{LQ} v_\Delta}^2$.  After imposing the \mue\ limit, the only coupling which is not well-constrained in the quasi-degenerate region is $Y_1^{21} = Y_1^{32}$, for which we take the largest allowed value of $\sim 10^{-1}~eV$.  If we take $M_{LQ}v_\Delta = 800~GeV~eV$ as in \cite{FileviezPerez:2008dw} ($v_\Delta$ is $\mathcal{O}\of{2~eV}$), then we find the total inclusive cross section for \etau\ is 0.08~$fb$ for 90~\gev~center-of-mass energy (well below the total inclusive cross section for this leptoquark calculated using the HERA limits).  With an integrated luminosity in excess of 10~\ifb, the EIC could be sensitive to \etau\ events mediated by the \stilde\ leptoquark.  Therefore, this $SU(5)$ GUT is one interesting example of a model which can satisfy existing \lfvem\ limits on both \mue\ conversion and \mueg\ (trivially, for the latter) but still give rise to a \lfvet\ signal that could be observable in the next generation of experiments, specifically \etau\ conversion at the EIC.

We conclude this section by noting that closer inspection of the \etau\ cross section reveals an interesting search prospect for this $SU(5)$ GUT leptoquark.  The \stilde\ leptoquark couples to down-type quarks, so the partonic level cross sections featuring a down quark in the initial state will dominate due to it from the proton p.d.f.s relative to other initial state quarks or antiquarks.  An initial-state down quark is only possible via the $u$-channel diagram for $F=0$ leptoquarks in \figref{fig:etaufeyndiagrams}; hence, the sub-processes proportional to $Y_1^{i 1}Y_1^{13}$ will give the largest contributions to the inclusive cross section.  Given the neutrino and \mue\ constraints discussed above, this product of the leptoquark couplings is maximized for $i = 3$, corresponding to the leptoquark coupling to an outgoing bottom quark.  Thus, in $e^-p$ collisions, an outgoing $\tau^-$ and a bottom quark jet may be a unique experimental signature of this $SU(5)$ GUT leptoquark.

\section{Conclusion}\label{sec:concl}

We have calculated cross sections for \etau\ processes as a function of the leptoquark ratios \lqratio\ for both scalar and vector leptoquarks; we have also calculated limits on these ratios, where applicable, from the most recent upper bound on \taueg\ decay for the scalar leptoquarks.  Our results, using an EIC center-of-mass energy equal to 90~\gev\ and 10~\ifb\ integrated luminosity, indicate that a search for charged lepton flavor violating \etau\ events is viable if such processes involve very massive ($M_{LQ}\gg\sqrt{s}$) leptoquarks at tree level; furthermore, new limits from the EIC in the event of a null result would be competitive with some of the present \taueg\ limits on quark flavor-diagonal combinations of scalar leptoquark couplings, though larger integrated luminosities and energies would be necessary to surpass all of the upper bounds imposed by \taueg.  On the other hand, an EIC search could substantially extend the reach into the space of quark off-diagonal couplings compared to current HERA and rare process limits. 

Additionally, we note that leptoquarks participate in strong interactions, and their pair production via gluon or quark/antiquark fusion can have large cross sections.  Searches for leptoquarks at the Tevatron and LHC have been considered \cite{Mitsou:2004hm,Eboli:1997fb,Belyaev:1998ki,Belyaev:2005ew,Kramer:2004df,FileviezPerez:2008dw}.  A search for \lfvet\ leptoquarks at the EIC would be complementary to these leptoquark searches at the Tevatron and LHC: the Tevatron and LHC are sensitive to the leptoquark mass and branching fractions, whereas \etau\ searches at the EIC depend on the ratio of the leptoquark couplings and mass, \lqratio.  Were the Tevatron or LHC to measure the mass of a leptoquark, the EIC could then provide information on its lepton flavor violating couplings to quarks and leptons.

While leptoquarks provide a framework in which \etau\ searches at the EIC could be feasible, there are many practical questions that are relevant for undertaking this \lfvet\ search.  For example, what processes would contribute to the backgrounds?  What would be the $\tau$ detection efficiency for the detectors?  These questions were previously answered for the HERA analyses, and their answers may provide some guidance although the EIC would be operating at a different energy and with different detectors.  Furthermore, it is important to consider the limits on \lfvet\ from rare processes.  From the study of \taueg\ decay in this analysis, it is clear that improvements in limits on other rare processes like $\tau\rightarrow 3e,\ \tau\rightarrow\pi e$, and decays of $K$ and $B$ mesons can impact the relevance of \etau\ searches at the EIC.  Although 10~\ifb\ integrated luminosity could be collected in a relatively short time at a high luminosity machine like the EIC, the time at which data collection could begin is uncertain and far in the future.  Thus,  the status of current and future experiments searching for \lfvet\ is necessary information when evaluating the prospective impact of a similar search at the EIC.  

Our analysis shows that leptoquark-induced \etau\ conversion at the EIC is worth further study, and the discussion of the $SU(5)$ GUT leptoquark in \secref{sec:neutrinos} shows that searches for \lfvet\ can still be relevant in light of the stronger limits on \lfvem\ processes.  Nonetheless, the leptoquark analysis presented here is only an initial study of \lfvet\ at the EIC.  There are important theoretical questions to be answered, such as, what are the \etau\ cross sections in other non-leptoquark models with \lfvet, and, can searches for \etau\ discriminate between various models?  These will be the subject of our future work. 

\section*{Acknowledgements}
The authors thank Abhay Deshpande, Pavel Fileviez P\'{e}rez, Krishna Kumar, and William Marciano for useful discussions.  MJRM also thanks the Aspen Center for Physics where part of this manuscript was completed. This work was supported in part by  U.S. Department of Energy contract DE-FG02-08ER41531 and by the Wisconsin Alumni Research Foundation.

\appendix 
\section{Scalar Leptoquark \taueg\ Results}\label{sec:appendix}
Here we show complete results for the coefficients $A_2^{L,R}$ contributing to the \taueg\ branching ratio.  All scalar leptoquarks are shown.  We have neglected the term proportional to $m_q^2/M_{LQ}^2$ in \eqnref{eq:lqA2}.  The charges of leptoquarks in an $SU(2)$ multiplet are differentiated using $Q_+, Q_0,$ and $Q_-$.
\begin{equation}\label{eq:allLQAs}
\begin{aligned}
&A_2^{\of{L,R}} = -\frac{1}{16\pi^2}\frac{N_c}{6}\of{\mathcal{Q}_q + \frac{\mathcal{Q}_{LQ}}{2}} \sum_{\alpha=1}^3\of{\frac{\lambda_{1\alpha}\lambda_{3\alpha}}{M_{LQ}^2}} \\
&\begin{aligned}
&\boxed{S_0^L} &\quad &\begin{aligned}&\mathcal{Q}_q = Q_{u^c} = -\frac{2}{3},\ \mathcal{Q}_{LQ} = \frac{1}{3} \\
&A_2^L = 0,\ A_2^R = \frac{1}{64\pi^2}\sum_{\alpha=1}^3\of{\frac{\lambda_{1\alpha}\lambda_{3\alpha}}{M_{LQ}^2}}\end{aligned}
\\
&\boxed{S_0^R} &\quad &\begin{aligned}&\mathcal{Q}_q = Q_{u^c} = -\frac{2}{3},\ \mathcal{Q}_{LQ} = \frac{1}{3} \\
&A_2^L = \frac{1}{64\pi^2}\sum_{\alpha=1}^3\of{\frac{\lambda_{1\alpha}\lambda_{3\alpha}}{M_{LQ}^2}},\ A_2^R = 0\end{aligned}
\\
&\boxed{\tilde{S}_0^R} &\quad &\begin{aligned}&\mathcal{Q}_q = Q_{d^c} = \frac{1}{3},\ \mathcal{Q}_{LQ} = \frac{4}{3} \\
&A_2^L = -\frac{1}{32\pi^2}\sum_{\alpha=1}^3\of{\frac{\lambda_{1\alpha}\lambda_{3\alpha}}{M_{LQ}^2}},\ A_2^R = 0\end{aligned}
\\
&\boxed{S_1^L} &\quad &\begin{aligned}&\mathcal{Q}_q = Q_{u^c} + 2Q_{d^c} = 0,\ \mathcal{Q}_{LQ} = Q_0 + 2Q_+ = 3 \\
&A_2^L = 0,\ A_2^R = -\frac{3}{64\pi^2}\sum_{\alpha=1}^3\of{\frac{\lambda_{1\alpha}\lambda_{3\alpha}}{M_{LQ}^2}}\end{aligned}
\\
&\boxed{S_{1/2}^L} &\quad &\begin{aligned}&\mathcal{Q}_q = Q_{u} = \frac{2}{3},\ \mathcal{Q}_{LQ} = Q_+ = \frac{5}{3} \\
&A_2^L = 0,\ A_2^R = -\frac{3}{64\pi^2}\sum_{\alpha=1}^3\of{\frac{\lambda_{1\alpha}\lambda_{3\alpha}}{M_{LQ}^2}}\end{aligned}
\\
&\boxed{S_{1/2}^R} &\quad &\begin{aligned}&\mathcal{Q}_q = Q_u + Q_d = \frac{1}{3},\ \mathcal{Q}_{LQ} = Q_{+} + Q_- = \frac{7}{3} \\
&A_2^L = -\frac{3}{64\pi^2}\sum_{\alpha=1}^3\of{\frac{\lambda_{1\alpha}\lambda_{3\alpha}}{M_{LQ}^2}},\ A_2^R = 0\end{aligned}
\\
&\boxed{\tilde{S}_{1/2}^L} &\quad &\begin{aligned}&\mathcal{Q}_q = Q_d = -\frac{1}{3},\ \mathcal{Q}_{LQ} = Q_+ = \frac{2}{3} \\
&A_2^L = 0,\ A_2^R = 0 \end{aligned}
\end{aligned}
\end{aligned}
\end{equation}

\bibliography{lq_lfv}

\end{document}